\documentclass[fleqn,usenatbib,useAMS]{mnras}

\usepackage{graphicx}	
\usepackage{epstopdf}

\usepackage{newtxtext,newtxmath, cuted, widetext}
\usepackage{float}
\usepackage[T1]{fontenc}

\usepackage{amsmath}	
\usepackage{color}
\usepackage{subcaption}
\newcommand{\totsamp}{1,743,152 }
\newcommand{\KMsamp}{256,381}

\title[Chemodynamical Ages in the Solar Neighborhood]{Chemodynamical Ages of Small-Scale Kinematic Structures of the Galactic Disc in the Solar Neighborhood from $\sim$250,000 K and M Dwarfs}

\author[I. Medan \& S. L\'{e}pine]{
    Ilija Medan,$^{1}$
	S\'{e}bastien L\'{e}pine,$^{1}$
	\\
	$^{1}$Department of Physics and Astronomy, Georgia State University, Atlanta, GA 30302, USA
}

\date{Accepted 2023 February 03. Received 2023 February 03; in original form 2022 September 24 }

\pubyear{2023}

\hypersetup{draft}

\begin{document}
\label{firstpage}
\pagerange{\pageref{firstpage}--\pageref{lastpage}}
\maketitle

\begin{abstract}
We combine photometric metallicities with astrometry from Gaia DR3 to examine the chemodynamic structure of $\sim$250,000 K dwarfs in the Solar Neighborhood (SN). In kinematics, we observe ridges/clumps of “kinematic groups”, like studies of more massive main-sequence stars. Here we note clear differences in both metallicity and vertical velocity as compared to the surrounding regions in velocity space and hypothesize this is due to differences in mean age. To test this, we develop a method to estimate the age distribution of sub-populations of stars. In this method, we use GALAH data to define probability distributions of W vs. [M/H] in age bins of 2 Gyr and determine optimal age distributions as the best fit weighted sum of these distributions. This process is then validated using the GALAH subset. We estimate the probable age distribution for regions in the kinematic plane, where we find significant sub-structure that is correlated with the kinematic groups. Most notably, we find an age gradient across the Hercules streams that is correlated with birth radius. Finally, we examine the bending and breathing modes as a function of age. From this, we observe potential hints of an increase in the bending amplitude with age, which will require further analysis in order to confirm it. This is one of the first studies to examine these chemodynamics in the SN using primarily low-mass stars and we hope these findings can better constrain dynamical models of the Milky Way due to the increase in resolution the sample size provides.
\end{abstract}


\section{Introduction}

Stars in the Milky Way are best described and categorized by considering their distributions in 3D-space, kinematics, abundances and age. With large, modern astronomical surveys, astronomers are now able to probe all of these parameters for statistically significant subsets of stars. Usually, all of these parameters are available in large numbers for giants, due to their brightness, which have allowed the study of kinematic groups in the Milky Way over large distances \citep[e.g.][]{ivezic2012,bovy2016, gaiadr2_kinematics, gaia_edr3_anticenter}. For low-mass stars however, it can be difficult to obtain such a rich multi-dimensional dataset. The most progress in building this ideal dataset for low-mass stars has been made in regards to 3D-space positions and kinematics, where Gaia \citep{gaiadr3} has provided the community with positions, parallaxes, proper motions and sometimes radial velocities for millions of low-mass stars in the Solar Neighborhood. 

Despite this advance, abundances for these stars have not been widely available, even from the largest spectroscopic surveys from the last decade, such as APOGEE \citep{apogeedr16}, GALAH \citep{galahdr3} and LAMOST \citep{lamost}, where abundances are only determined for a subset multiple orders of magnitude smaller than the Gaia dataset. This has however improved with Gaia DR3, which includes stellar parameters for 100s of millions of stars based either on the high resolution RVS spectra \citep{gaiadr3_RVS} or the low resolution BP/RP spectra \citep{gaiadr3_BPRP}. The metallicities derived from BP/RP spectra have been found to be dominated by large systematic errors and have been advised to only be used in a qualitative sense \citep{gaiadr3_BPRP}. Stellar parameters derived from the higher resolution RVS spectra are more accurate but only available for a smaller subset of relatively brighter stars; in addition, as will be shown later in this work, they also appear to suffer from systematic errors for lower-mass stars. Finally, photometry has also been used to estimate metallicities of stars in Gaia eDR3. For example, \citet{Xu2022} recently calibrated a photometric metallicity relationship applicable to 27 million FGK stars. This relationship is however not applicable to dwarfs less massive than about K5. In our recent work \citep{medan2021}, we have calibrated an improved photometric metallicity relationship that expands this range to subtype M3V so the metallicities for lower-mass stars can be estimated.

Beyond this, age determinations of low-mass field stars has remained elusive despite recent progress in gyrochronology \citep{angus2019}, which seek to provide adequate age determinations from intensive photometric monitoring, most reliably for younger ($<1$ Gyr) stars. Progress has also been made recently again with the release of Gaia DR3, from which ages can be better estimated from model isochrones thanks to the Gaia RVS spectroscopic parameters, which can be used as inputs \citep{Kordopatis2022}. In any case, isochrone ages remain most accurate for stars near or above the main-sequence turnoff, and errors are much larger for lower-mass main-sequence stars.

The current lack of reliable stellar parameters for low-mass stars in the Solar Neighborhood is disappointing, as a dense map of the local chemodynamical structure would be a great complement to the large but spatially sparse map drawn by the Milky Way giants. Low-mass stars in the Solar Neighborhood provide a unique opportunity to study small-scale structures in kinematic space that are blurred out over large distances in the giants datasets. These small-scale structures were first seen in great detail by \cite{nordstrom2004}, where ridges in the kinematic space distribution were observed. Unlike open clusters, which also manifest as clumps in kinematic space, \cite{nordstrom2004} demonstrated that these steams had fairly large spreads in metallicity and age, suggesting a dynamical origin of these streams rather than the dissolution of star clusters. Specifically, it has be theorized that non-axisymmetric perturbations from spiral arms, bars, etc. can cause such groupings in velocity space in the Solar Neighborhood  \citep[e.g.][]{dehnen2000, fux2001, famaey2005, quillen2005, bovy2010_2, hunt2018, wojno2018}.

The evidence for such origins has also grown over the past couple of decades with some limited data sets that have become available. For example, \cite{bovy2010} demonstrated that the kinematic members of some of these "streams" have metallicity distributions that differ from that of the general field population, which is expected for stars with a distinct dynamical origin. Also, based solely on the kinematics of these groups, \citet{quillen2018} demonstrated that the ridges formed in  kinematic space are consistent with orbits that would have recently interacted with nearby spiral arms. Additionally, these kinematic structures can be successfully replicated in computer simulations of the Milky Way with various non-axisymmetric potentials (i.e. bar, spiral arms) which supports the idea of orbital resonances as a primary cause \citep[e.g.][]{Fragkoudi2019, hunt2019, barros2020}. It remains unclear however which parameters/properties of the non-axisymmetric potentials can best explain the observed kinematics of local stars, as multiple combinations of potentials and parameters can reproduce the current set of observations.

One way to break down these degeneracies is to probe local stars in more dimensions than 3D kinematics. Ideally, a full understanding of the kinematic-age-metallicity distribution of local stars may better constrain models of the Milky Way potential. While this has been attempted for more massive main sequence stars \citep[e.g.][]{nordstrom2004, antoja2008}, this has not been done for low-mass K and M dwarfs. In this study, we will examine the kinematic-age-metallicity distribution of local K and M dwarfs, which due to the larger number of these stars compared to Solar-type stars in the Solar Neighborhood, allows us to examine smaller-scales trends in this parameter space than in previous studies. Due to the lack of adequate age determinations for these low-mass stars though, we will calibrate and validate a method to determine probable age distributions of stars based on metallicity and kinematics. The resulting chemodynamic age distributions shown here agrees with those found in previous studies at larger scales, and at smaller scales reveal relations to kinematics that may give further insight into the timescale of potential features (i.e. spiral arms and bar) and to the general merger history of the disk. Most importantly, this study will work as a framework for how such a chemodynamical study can be conducted with low-mass stars, such that when future surveys, like SDSS-V, become more prevalent, this analysis may be expanded to other volumes near the Solar Neighborhood.

A brief outline of this study is as follows. In Section \ref{sec:data} we describe how the photometric metallicites for the low-mass stars are acquired and how we identify the kinematic streams in the Solar Neighborhood, including the identification of several sub-streams. In Section \ref{sec:var_x_U} we discuss the variations in metallicity and kinematics that are observed in these groups, and the possible origin of such variations. In Section \ref{sec:age_cal} we describe a methodology for determining chemodynamical age distributions for groups of stars and assess the validity of this method. In Section \ref{sec:discuss} we discuss the resulting age distributions resulting from our chemodynamical method, both in the context of past studies and new features that we observe. Additonally we examine the bending and breathing modes in the kinematic plane and how they relate to age. Finally, in Section \ref{sec:conclusion} we summarize the observations in this paper and discuss the impact these observations could have on future studies of the origin of these kinematic structures.

\section{Data}\label{sec:data}

\subsection{Photometric Metallicites}

\begin{figure*}
	\includegraphics[width=\textwidth]{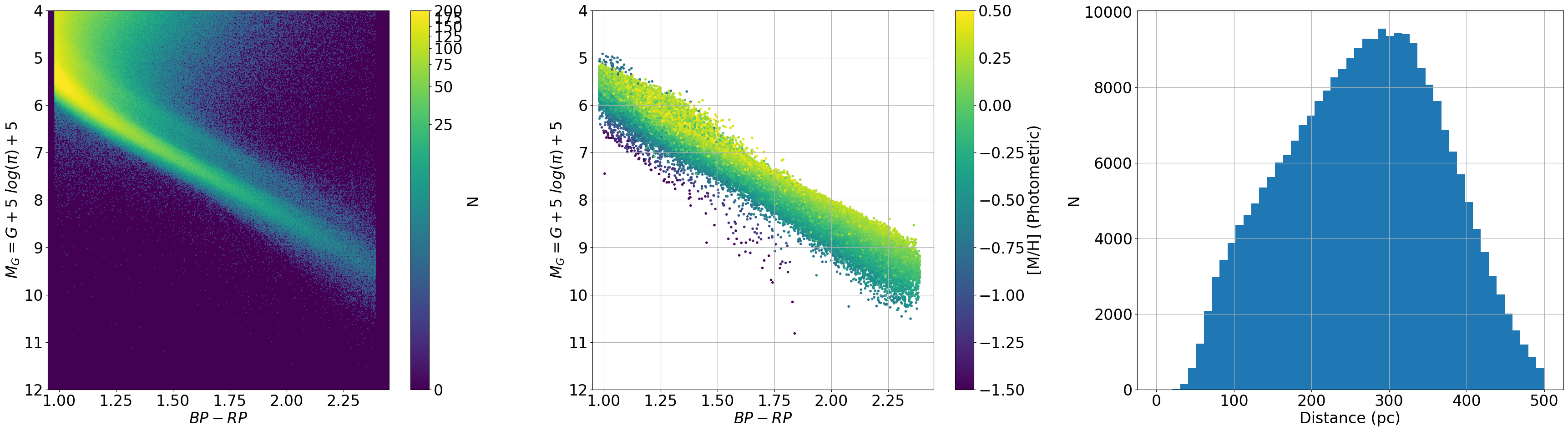}
	\caption{\textit{Gaia} HR diagram (left and middle panels) and histogram of distances for various stellar samples. The left panel shows the HR diagram for \totsamp stars in \textit{Gaia} that are with $G < 14$ and $RP < 14$, and also have colors consistent with a K/M stars \citep{mamajek2013}. The middle and right panels include the subsample of \KMsamp K and early M dwarfs with $G < 14$ and $RP < 14$, and have reliable photometric metallicity estimates using the calibrated relationship from \citet{medan2021}. The large discrepancy in numbers between the two samples is mostly due to the large number of stars with saturated magnitudes in Pan-STARRS, which must be excluded. In the HR diagram in the middle panel, the data points are colored according to their estimated photometric metallicity.}
	\label{fig:photo_metals}
\end{figure*}

From Gaia DR3, we select stars with $G < 14$ and $RP < 14$ (as they should have radial velocity measurements), have a color consistent with a K or early M star \citep[$0.98<B_P-R_P<2.39$;][]{mamajek2013}, and have $M_G>4$. These data can be selected from the Gaia archive by the query:
\begin{verbatim}
	SELECT *
	FROM gaiadr3.gaia_source as g3
	WHERE g3.phot_g_mean_mag <= 14 AND 
	             g3.phot_rp_mean_mag <= 14 AND
	             g3.phot_bp_mean_mag - 
	             g3.phot_rp_mean_mag > 0.98 AND 
	             g3.phot_bp_mean_mag - 
	             g3.phot_rp_mean_mag < 2.39 AND 
	             g3.parallax > 0 AND 
	             g3.phot_g_mean_mag + 
	             5 * log10(0.001 * g3.parallax) + 5 > 4
\end{verbatim}
This results in a sample of \totsamp stars that are shown in the left panel of Figure \ref{fig:photo_metals}. To obtain additional optical and infrared photometry, we cross-match these stars with the Two-Micron Sky Survey \citep[2MASS;][]{2mass}, AllWISE \citep{allwise} and the Panoramic Survey Telescope and Rapid Response System \citep[Pan-STARRS;][]{ps1} using the Bayesian cross-macthing method from \citet{medan2021}. To significantly clean this sample, and estimate very accurate photometric metallicites, we remove all stars with $\sigma_\pi / \pi > 0.2$, stars with RUWE $> 1.4$ to ensure well behaved astrometric solutions \citep{lindengren2021}, stars that don't have the required photometry from all surveys and stars with $g_{PS1}<13.5$, as it is likely that brighter objects have saturation issues in Pan-STARRS. This removes a large number of sources, reducing the sample to only 508,337 stars. Before estimating photometric metallicities for these, we correct all photometry for extinction using the 3D dust map from \citet{baystar19}. Values of $A_\lambda$ are calculated for Pan-STARRS, assuming $R_V=3.1$, using the results from \citet{schlafly2011}, and $A_\lambda$ values for 2MASS and AllWISE using the results from \cite{davenport2014}. These values of $A_\lambda$ are used to correct all photometric measurements for extinction. 

We then apply the calibrated photometric metallicity relationship from \citet{medan2021} to estimate metallicity values for K and early M dwarfs, and remove stars that appear overluminous and are likely to be unresolved pairs, as discussed in \citet{medan2021}. A full discussion of the method to derive the photometric metallicity relationship can be found in  \citet{medan2021}, but we also include a brief summary here. The relationship between photometry, absolute magnitudes, and metallicity was calibrated using a Gaussian Process Regressor with a radial basis function (RBF) kernel and a white-noise kernel proportional to the average error of the derived abundances, in an iterative manner. For this process, the training set used stars with derived stellar parameters from APOGEE spectra. In the first iteration of the calibration, all possible combinations of 2MASS, AllWISE and Pan-STARRS colors and absolute magnitudes were used as inputs. Then each color/absolute magnitude was removed in turn and the calibration re-evaluated. The color or absolute magnitude producing the smallest change in the mean squared error was then omitted for the remaining iterations. This continues until two inputs remained. The optimal inputs for the regressor were determined from the combination of colors/absolute magnitudes that minimized the mean squared error throughout this entire process. In the next step, unresolved binaries were removed from the training sample. This was done by using the best combination of color and absolute magnitude from the previous step. In this magnitude vs. color-space, overluminous stars, relative to other stars of similar metallicity, can be identified and removed removed from the training sample. With this cleaned sample, new optimal inputs were found based on the process explained in the first step. These provide the inputs for the final calibrated relationship, where the final, optimal inputs were found to be: $M_g$, $g-y$, $y-W2$, $J-W2$, and $W1-W2$.

The resulting photometric metallicities do have the limitation of having larger uncertainties at lower metallicities ($[M/H] < -1.5$) and at the lower-mass end of the calibration range, with the calibration valid only over the color range $0.98 < BP-RP< 2.39$. To account for this, we finally restrict the sample to stars with photometric metallicities that have an uncertainty less than 0.3 dex at the 95\% confidence level. This results in the final sample of \KMsamp stars shown in the center panel of Figure \ref{fig:photo_metals} and listed in Table \ref{tab:sample_table}, where $\sim84\%$ of the sample has an uncertainty on the photometric metallicity estimate less than 0.12 dex at the 95\% confidence level. We find that most of these stars are within a few hundred parsecs of the Sun (right panel of Figure \ref{fig:photo_metals}), confirming that we are probing a very local population of low-mass stars in the Solar Neighborhood.

\begin{table*}
	\scriptsize
	\centering
	\caption{Catalog of the \KMsamp K and early M dwarfs from Gaia DR3 stars with $G < 14$ and $RP < 14$, and with reliable photometric metallicity estimates from the calibrated relationship from \citet{medan2021}. The table includes the Gaia DR3 information, estimated photometric metallicity and kinematic information used in this study.}
	\label{tab:sample_table}
	\begin{tabular}{ccccccccc}
		\hline
		Gaia DR3 Source ID & $\alpha$ & $\delta$ & [M/H] & $\sigma_{[M/H]}$ & $U$ & $V$ & $W$ & $x_{mix}=R_G \ cos(\Phi)$ \\
		&  [deg] & [deg] & [dex] & [dex] & [km/s] & [km/s] & [km/s] & [kpc]  \\
		\hline
		4282728430212942464 & 283.19308398639   & 6.06158030747 & $-$0.41 & 0.06    & 21.902   & $-$21.318    & 27.192     & 7.746\\
		4282091812993298816 & 282.17970392886   & 4.51928207912 & 0.11 & 0.08    & $-$4.974   & $-$13.194    & $-$4.150    &  7.838\\
		4128312325715430272 & 258.23667888083 & $-$19.43635165157 & $-$0.05 & 0.06   &  $-$5.982    &  8.821  &  $-$39.739   &   8.635\\
		4281805420263719040 & 282.59675875222   & 3.81918889623 & $-$0.00 & 0.06   & $-$30.739  &  $-$51.144  &   $-$9.298  &    6.736\\
		4128313841869775872 & 258.21227674779 & $-$19.38116504155 & $-$0.31 & 0.07  &  $-$26.217  &  $-$16.285  &  $-$14.148  &    7.736\\
		4128313871904058112 & 258.22275834911 & $-$19.37048145137 & 0.23 & 0.06  &  $-$29.797  &    4.478  &  $-$12.395   &   8.523\\
		4282963175964286336 & 278.66928326503   & 2.79886648525 & 0.17 & 0.06 &   $-$10.806 &   $-$3.267  &  $-$24.859    &  8.245\\
		4128257625010658432 & 257.93797919361 & $-$19.96636803101 & $-$0.06 & 0.06  &   $-$6.272     & 6.238  &  $-$38.014  &    8.619\\
		4282746091118405376 & 283.05401725726   & 6.21708359404 & 0.14&  0.09  &  102.050  &  $-$39.787   &  $-$3.825 &     6.931\\
		4282370848412579712 & 284.23740869742   & 5.02662499513 & 0.46&  0.07   &   6.946 &   $-$24.127  &  $-$10.468 &     7.487\\
		\hline
	\end{tabular}
	\\
	NOTE -- This table is published in its entirety in a machine-readable format. A portion is shown here for guidance regarding its form and content.
\end{table*}

To verify the precision of our photometric metallicity estimates, we compare our photometric metallicities to results from APOGEE DR14 \citep{apogeedr14}, APOGEE DR17 \citep{apogeedr17}, GALAH DR3 \citep{galahdr3}, and both the BP/RP \citep[i.e. GSP-Phot;][]{gaiadr3_BPRP} and RVS \citep[i.e. GSP-Spec;][]{gaiadr3_RVS} derived metallicities from Gaia DR3. Stars between our sample and the spectroscopic catalogs are matched by Gaia identifier, with GALAH DR3 providing DR3 identifiers and APOGEE DR2 identifiers. For APOGEE, we match our sample to DR3 using the table \texttt{gaiadr3.dr2\_neighbourhood} from the Gaia archive and consider the best match to be the DR2 source with the minimum angular distance from the DR3 source. For the Gaia DR3 GSP-Phot and GSP-Spec metallicites, there are known systematic errors in the values listed in the Gaia archive. Because of this, \citet{gaiadr3_BPRP} offers a Python package\footnote{\url{https://github.com/mpi-astronomy/gdr3apcal}} with empirical calibrations models of the stellar parameters from GSP-Phot and \cite{gaiadr3_RVS} provides a metallicity calibration as a function of $log(g)$ that is based on comparing values to APOGEE, GALAH and RAVE. We use both of these corrections in the below comparison.

\begin{figure*}
	\includegraphics[width=0.82\textwidth]{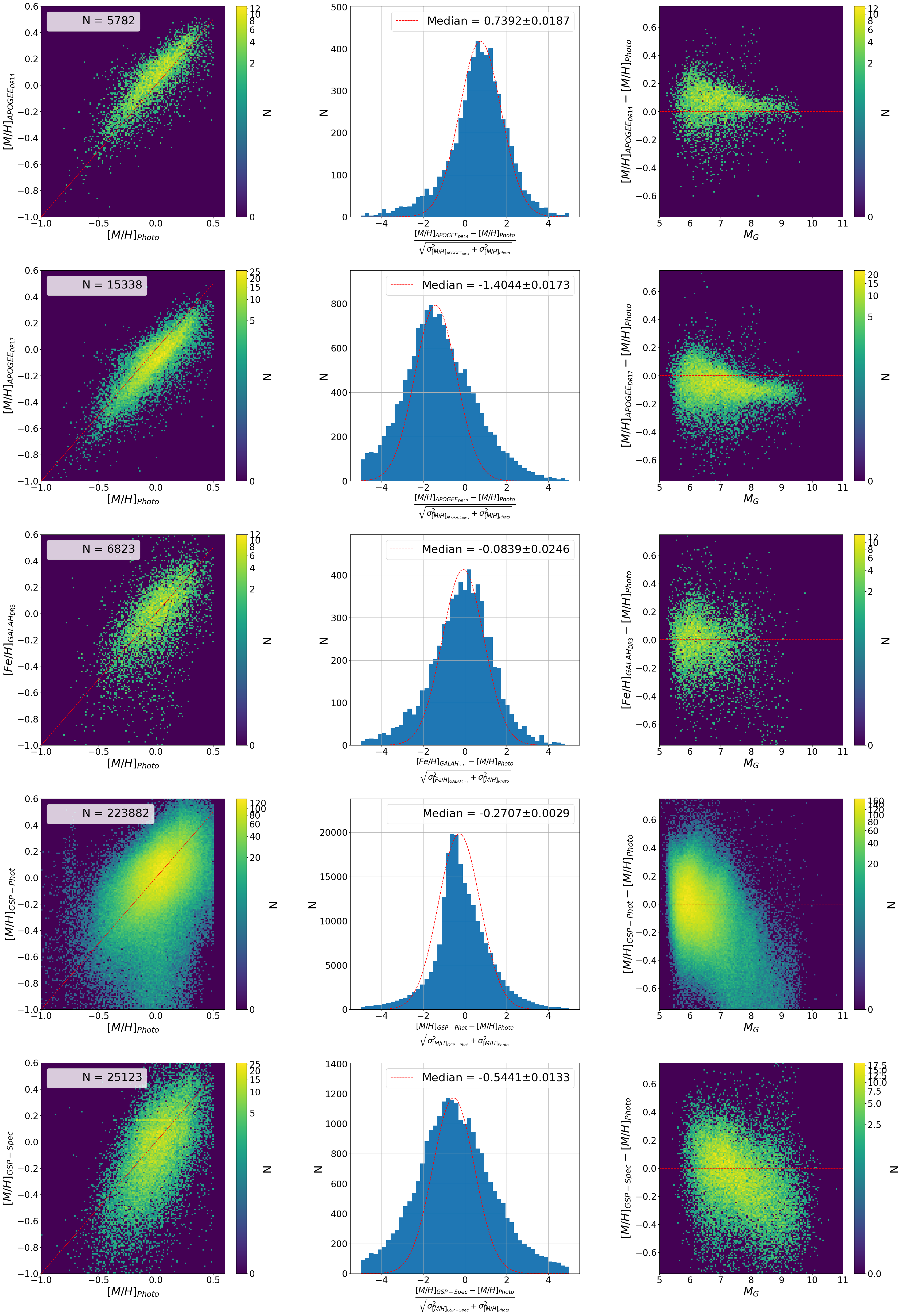}
	\caption{Comparison between our estimated photometric metallicties and spectroscopic metallicities from APOGEE DR 14 (first row), APOGEE DR17 (second row), GALAH DR3 (third row), GSP-Phot corrected metallcities from Gaia BP/RP spectra (fourth row) and GSP-Spec corrected metallicities from Gaia RVS spectra. The left column shows the spectroscopic metallicity vs. the photometric metallicity. The middle column shows the difference between the spectroscopic and photometric metallicity divided by the total error on the difference between the two. The median of the distribution provides the relative accuracy of our estimates and is shown in the legend. The red dashed line shows a normal distribution with mean equal to this median and standard deviation of one, where any deviation at the tails can be indicative of systematic offsets in one or both measurements. The right column shows the difference in metallicity as a function of absolute magnitude of the star. While the photometric estimates show consistency with APOGEE and GALAH values, we see systematic differences in both the Gaia BP/RP (GSP-Phot; fourth row) and RVS (GSP-Spec; fifth row).}
	\label{fig:compare_photo_metals}
\end{figure*}

Figure \ref{fig:compare_photo_metals} shows the comparison of metallicities between this study ($[M/H]_{photo}$) and the values from each of the spectroscopic surveys listed above. The central column of Figure \ref{fig:compare_photo_metals} shows the difference between the spectroscopic and photometric metallicity divided by the total error on the difference between the two, where the median of the distribution provides the relative accuracy of our photometric estimates. The red dashed line is a normal distribution with mean equal to this median and standard deviation of one, where any deviation at the tails can be indicative of systematic offsets in one or both measurements. The comparison to APOGEE DR14 and GALAH DR3 mostly follows this $1\sigma$ distribution. This makes sense for the former as the photometric metallicities were calibrated with this version of APOGEE metallicities. The good agreement with GALAH DR3 will be crucial for subsequent analyses in this paper that rely on metallicities from this survey. While systematic offsets are noted, the accuracy of our estimates can mostly ignored as we are only analyzing relative differences in the chemodynamical sub-structure. When comparing our results to those from spectroscopic studies however, these offsets should be taken into account.

While our photometric metallicity estimates are generally consistent with the APOGEE and GALAH values, we note systematic differences with the Gaia-based metallicity values. This is shown in the right column of Figure \ref{fig:compare_photo_metals}, where we examine the difference in metallicity as a function of absolute magnitude of the star. For both the Gaia BP/RP (GSP-Phot; fourth row) and RVS (GSP-Spec; fifth row) metallicities  we see that there is a dependence on the difference in metallicity with absolute magnitude (which is a proxy for mass) that is not present when comparing to the other spectroscopic surveys. For both cases, we see that the metallicities are systematically underestimated for lower-mass K and early M dwarfs. This reinforces the idea that Gaia-derived metallicity values for local, low-mass stars should be used with caution and that more calibration is still needed for the lowest mass Gaia stars.

\subsection{Kinematic Groups}\label{sec:kin_groups}

\begin{figure*}
	\includegraphics[width=\textwidth]{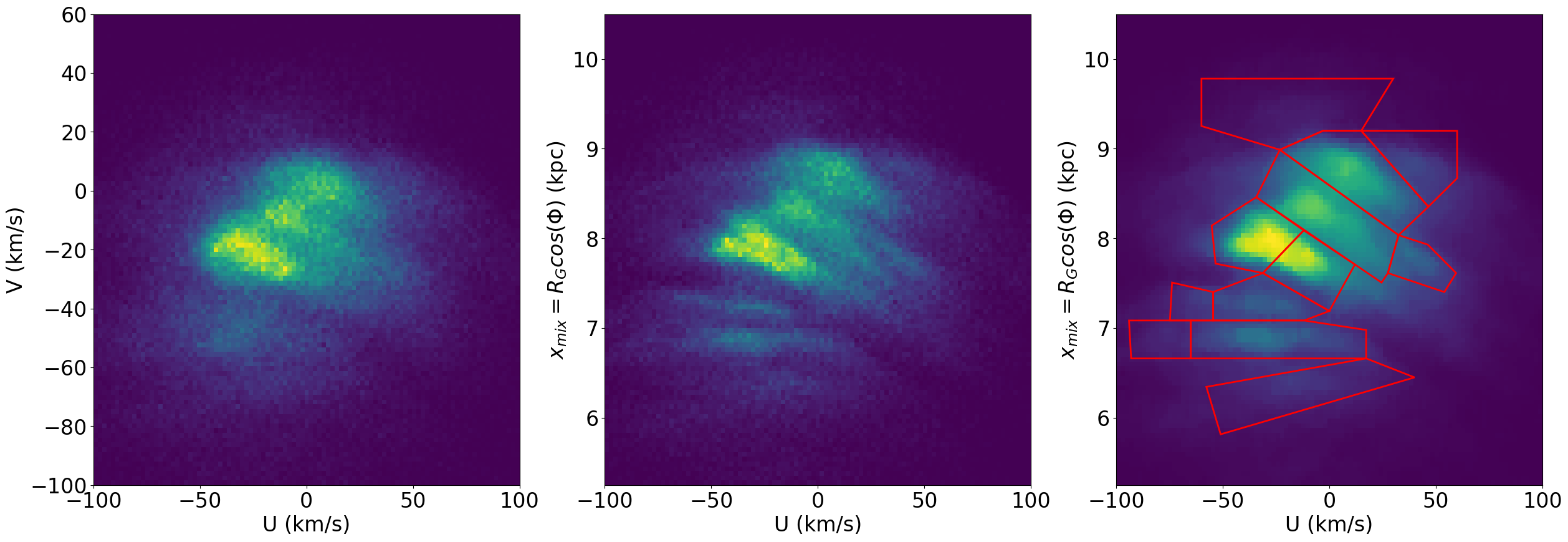}
	\caption{$V$ vs $U$ (left panel) and $x_{mix}=R_G cos(\Phi)$ vs $U$ (center and right panels) for \KMsamp K and early M dwarfs with $G < 14$ and $RP < 14$, and have reliable photometric metallicity estimates using the calibrated relationship from \citet{medan2021}. In the above, $R_G=\left[R\left(V+v_\odot\right)\right]/v_{LSR}$ and $\Phi=sin^{-1}\left[d \times sin\left(l\right)/R\right]$, where $R$  is the distance from the Galactic center in the $xy$ plane; $U$ is the velocity towards the Galactic anticenter; $V$ is the velocity in the direction of Galactic rotation; $v_\odot=248.5$ km/s \citep{hunt2020}, and is the Sun's velocity in the direction of Galactic rotation; $v_{LSR} = 235$ km/s and is the local standard of rest velocity; $d$ is the distance from the Sun; and $l$ is the Galactic longitude. The right panel is a 2D histogram of the data in the center panel that has been smoothed using a Gaussian kernel with a size of $\sigma=1.25$ pixels. The red polygons shown in the right figure are drawn by hand in order to provide a first guess of the kinematic groups that show up as over densities int he above plots.}
	\label{fig:UV_vs_xmix}
\end{figure*}

To examine the kinematic groups present in our photometric metallicity sample, we transform \textit{Gaia} 6D astrometric data into Galactic cartesian coordinates: Galactic radius ($R$), Galactic height ($z$), velocity towards the Galactic anti-center ($U$), velocity in the direction of Galactic rotation ($V$) and velocity in the direction perpendicular to the Galactic plane ($W$). For this transformation we assume $R_\odot\approx8.1$ kpc \citep{grav_constant_2019} and $z\approx21$ pc \citep{bennett2019}. The left panel of Figure \ref{fig:UV_vs_xmix} shows the resulting $V$ vs $U$ kinematic distribution for our sample, where clumps in velocity space are prominent. These clumps correspond to known kinematic groups, or "streams", in the Solar Neighborhood and are similar to what has been observed previously with Gaia DR2 for more massive stars \citep{gaiadr2_kinematics}.

To more clearly identify these clumps in orbital space, we transform some of these parameters to examine the clumps in a mixed coordinate system. Specifically, we follow the coordinate transformation outlined by \citet{hunt2020}. For this transformation, orbits in the disk which are nearly circular and occuring in a potential that is axisymmetric, can be given an epicyclic approximation. For this approximation, it is assumed that the motion of a  star with some angular momentum can be split into the mean motion of the guiding center of the star superimposed with its epicyclic oscillation around this mean. This guiding center is defined as the circular radius of an orbit that would have the same angular momentum of the star:
\begin{equation}
	R_G=\frac{L_z}{v_{LSR}}=\frac{R\left(V+v_\odot\right)}{v_{LSR}}
\end{equation}
where we assume $v_\odot = 248.5$ km/s, where this is calculated from $R_\odot$ and the proper motion measurement of Sgr A$_*$ \citep{reid2020} as was done in \citet{hunt2020}, and $v_{LSR} = 235$ km/s, to match the value from \cite{frankel2020}. To account for the current azimuth $(\Phi)$ of the star, our mixed coordinate is then:
\begin{equation}
	x_{mix}=R_G \ cos(\Phi)
\end{equation}
This mixed coordinate, while not fully describing the orbit as with action coordinates, still does combine the azimuthal action with the physical location of the star to give some indication of the orbital label for a star. This should allow for stars on similar orbits to be better grouped together compared to a more simple UV diagram.

The middle and right panels of Figure \ref{fig:UV_vs_xmix} show our modified kinematic diagram with the $V$ velocity replaced by our mixed coordinate. In the $x_{mix}$ vs. $U$ plane, the clumps have higher contrast than in the $V$ vs. $U$ plane (Figure \ref{fig:UV_vs_xmix} , left panel). Additionally, the gaps between kinematic groups are better defined, especially for the large curved gap located around $(U,x_{mix})=(-25,7.5)$. The reason for this is that, as we are probing a relatively small volume close to the Sun, $cos(\Phi)\approx 1$, which implies that $x_{mix}\propto L_z$. If these kinematic groups are stars in a shared orbit, it would then make sense why they would have similar angular momenta and become well separated in a plane that incorporates this aspect of the orbit. We use this mixed coordinate system to identify kinematic groups in our sample.

To identify groups in kinematic space, many contemporary studies have utilized a wavelet based analysis to identify peaks in the kinematic distribution \citep[e.g.][]{gaiadr2_kinematics, ramos2018}. In this study, we are not as concerned with discovering new structures in our data, but are more interested in roughly identifying past kinematic groups in our mixed coordinate system. Because of this, we will be attempting to visualize the most prominent kinematic groups, and not all of the small detections that may be able to be identified as with a wavelet analysis. While not as statistically rigorous, this rough identification of previous groups will still allow us to compare the global trends observed here to those from previous studies.

\begin{figure*}
	\centering
	\includegraphics[width=0.9\textwidth]{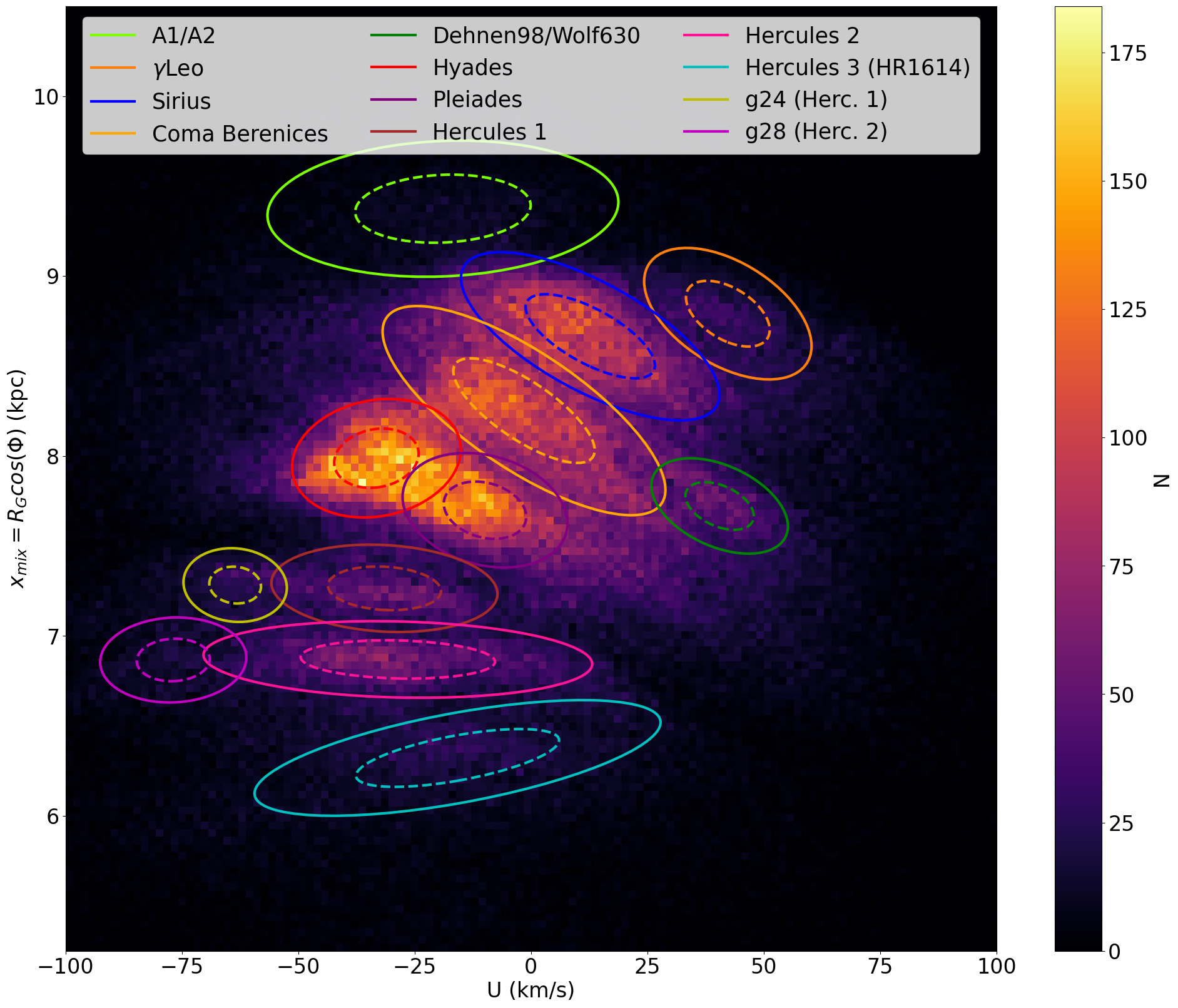}
	\caption{$x_{mix}=R_G cos(\Phi)$ vs $U$ for \KMsamp K and early M dwarfs with $G < 14$ and $RP < 14$, and with reliable photometric metallicity estimates based on the calibrated relationship from \citet{medan2021}. In the above, $R_G=\left[R\left(V+v_\odot\right)\right]/v_{LSR}$ and $\Phi=sin^{-1}\left[d \times sin\left(l\right)/R\right]$, where $R$ is the distance from the Galactic center in the $xy$ plane; $U$ is the velocity towards the Galactic anticenter; $V$ is the velocity in the direction of Galactic rotation; $v_\odot=248.5$ km/s \citep{hunt2020} and is the Sun's velocity in the direction of Galactic rotation; $v_{LSR} = 235$ km/s and is the local standard of rest velocity; $d$ is the distance from the Sun; and $l$ is the Galactic longitude. The ellipses in the plot correspond to the 1$\sigma$ (dashed lines) and 2$\sigma$ (solid lines) regions around the median for each kinematic group, as determined by a PCA analysis. The medians and standard deviations of each kinematic group are listed in Table \ref{tab:group_ellipses}.}
	\label{fig:PCA_ellipses}
\end{figure*}

As a first step in identifying these groups, we create a 2D histogram of the data in the $x_{mix}$ vs. $U$ plane that is then smoothed using a Gaussian kernel with a size of $\sigma=1.25$ pixels. This reduces the noise and allows us to visually identify regions of interest.Visually we draw polygon regions around clumps that look well defined and independent in this smoother 2D histogram, and that are commonly identefied in wavelet analyeses of more massive stars in contemporary studies These polygons are shown in the right panel of Figure \ref{fig:UV_vs_xmix}. Using the data contained within each region, we perform a PCA decomposition, implemented in \textit{scikit-learn} \citep{sklearn}, in order to determine the principal axis, median and standard deviation in either direction for the data within that polygon. This allows for each kinematic group to be visualized as an ellipse in the $x_{mix}$ vs. $U$ plane, where the size of the ellipse is defined as some number of standard deviations from the center. This is illustrated in Figure \ref{fig:PCA_ellipses} for each of the polygon regions, where the medians and standard deviations of each kinematic group are listed in Table \ref{tab:group_ellipses}. 

These kinematic groups are well associated with the groups most commonly found in the literature. We do note that the extent/centers are not necessarily the same as past studies, but the overall structures as well associated with those in the literature. A brief review of these groups are as follows:
\begin{itemize}
	\item \textbf{A1/A2:} A high angular momentum group(s) identified by both \citet{gaiadr2} and \citet{ramos2018}. In these studies, the wavelet analyses identified them as two independent structure, but due to the smaller volume probed in the study we observe much fewer stars in raw counts here and merge them into one group for visual purposes.
	\item \textbf{$\gamma$Leo:} Corresponds to a structure referred to as $\gamma$Leo in e.g.~\citet{antoja2012}. In past studies, it has been unclear if this is a continuation of the Sirius stream as they both share similar angular momenta \citep{Kushniruk2019}. In the mixed coordinate system used here they seem well separated though, so we identify it as its own structure here.
	\item \textbf{Sirius}: A very common kinematic group that has been identified in most studies related to kinematic structures in the Solar Neighborhood.
	\item \textbf{Coma Berenices:} A kinematic group that, while appearing as singular clump in kinematic space, has been subdivided in past studies due differences in the spatial distribution for negative $U$ portion of the group \citep{monari2018, quillen2018b}. This incomplete phase mixing will be discussed later in this study.
	\item \textbf{Dehnen98/Wolf630:} Two historical groups next to Coma Berenices that were initially indetified by \citet{dehnen1998} and \citet{eggen1971}, respectively. Similar to A1/A2, due to lower counts we treat them as one group for visual purposed in this study.
	\item \textbf{Hyades/Pleiades:} Two kinematic groups that roughly coincide with the velocity of the open clusters of the same names. Here we select  Hyades as the low $U$ structure and Pleiades as the high $U$ structure as this is what has been done historically \citep[e.g.][]{antoja2012}. We do note here that the Hyades does appear as two arches in our mixed coordinate system, indicating the structure may be more complex than this.
	\item \textbf{Hercules 1/ Hercules 2:} A very common kinematic group with low angular momentum that has been separated into two branches  \citep[e.g.][]{antoja2012}.
	\item \textbf{Hercules 3 (HR1614):} A kinematic group at very low angular momentum that was identified based on being quite metal rich \citep{Feltzing2000, desilva2007}. From more recent N-body simulations, it has been shown that we expect the Hercules stream to manifest as a trimodal structure \citep{asano2020}. This is consistent with the observed structure in Gaia DR2 \citep{gaiadr2_kinematics}, which is even more prominent in this study, so we choose to mainly refer to this structure as Hercules 3, but reference HR1614 for historical purposes.
	\item \textbf{g24 (Herc. 1)/g28 (Herc. 2):} Two small sub-streams that appear independent from Hercules 1 and 2, but still have similar angular momenta. These two groups were previously identified by \citet{Kushniruk2019}, which is where the names g24 and g28 come from.
\end{itemize}

\begin{table*}
	\centering
	\caption{Median and standard deviations for each kinematic group in the $x_{mix}$ vs $U$ plane. The standard deviations are along a coordinate frame rotated by the angle $\theta$ in the clockwise direction. The last two columns of the table give the number of stars within some standard deviation from the mean for each kinematic group.}
	\label{tab:group_ellipses}
	\begin{tabular}{lccccccc} 
		\hline
		Kinematic Group & $U_{median}$ & $x_{mix,median}$ & $\sigma_U$ & $\sigma_{x_{mix}}$ & $\theta$ & $\sigma<1\sigma$ & $1\sigma<\sigma<2\sigma$\\
		 & [km/s] & [kpc] & [km/s] & [kpc] & [$^\circ$] & & \\
		\hline		
		A1/A2 & -18.920 & 9.374 & 56.542 & 0.565 & -0.057 & 1744 & 4229 \\
		$\gamma$Leo & 42.295 & 8.791 & 26.975 & 0.477 & 0.573 & 1790 & 4032 \\
		Sirius & 12.737 & 8.665 & 41.664 & 0.505 & 0.668 & 10006 & 19849 \\
		Coma Berenices & -1.500 & 8.252 & 45.580 & 0.569 & 0.830 & 12850 & 27801 \\
		Dehnen98/Wolf630 & 40.520 & 7.722 & 22.019 & 0.360 & 0.441 & 1907 & 4458 \\
		Hyades & -33.207 & 7.988 & 27.189 & 0.488 & -0.155 & 9246 & 19834 \\
		Pleiades & -9.877 & 7.699 & 26.569 & 0.464 & 0.238 & 7597 & 16079 \\
		Hercules 1 & -31.487 & 7.265 & 36.434 & 0.361 & 0.072 & 3021 & 6179 \\
		Hercules 2 & -28.604 & 6.869 & 62.638 & 0.317 & 0.035 & 5227 & 9077 \\
		Hercules 3 (HR1614) & -15.782 & 6.322 & 65.429 & 0.380 & -0.260 & 2652 & 5565 \\
		g24 (Herc. 1) & -63.559 & 7.283 & 16.659 & 0.306 & 0.080 & 693 & 1682 \\
		g28 (Herc. 2) & -76.853 & 6.867 & 23.553 & 0.354 & -0.046 & 699 & 1853 \\
		\hline
	\end{tabular}
\end{table*}

\section{Results}

\subsection{Variations in the $x_{mix}$ vs. $U$ Plane}\label{sec:var_x_U}

\begin{figure*}
	\includegraphics[width=\textwidth]{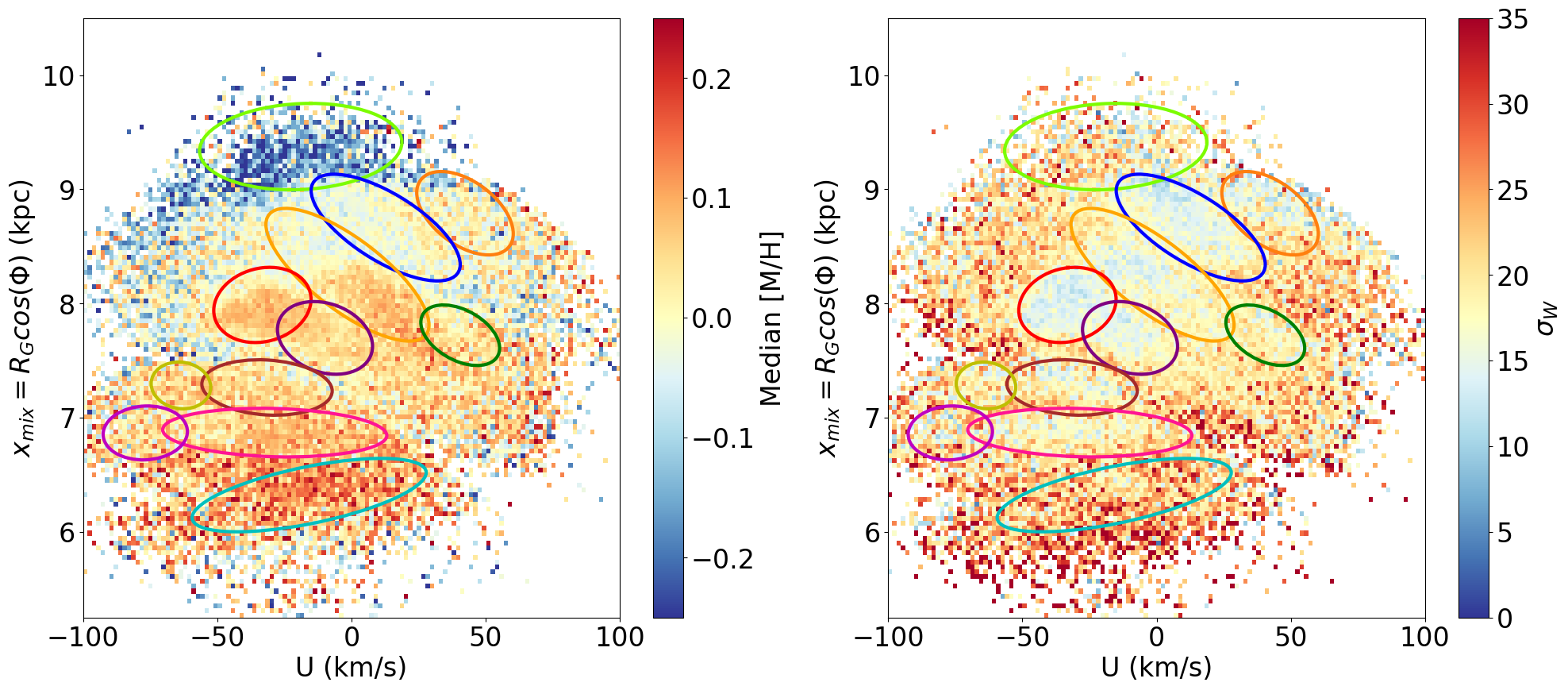}
	\caption{Maps of median [M/H] (left panel) and $\sigma_W$ (right panel) in the $x_{mix}$ vs. $U$ plane. The ellipses for each stream at the $2\sigma$ level are shown for reference. In both panels, the median or standard deviation of a pixel in the grid is only calculated for pixels with $N>5$.}
	\label{fig:metal_W_UV_plane}
\end{figure*}

In Section \ref{sec:kin_groups}, we identified a number of streams in the Solar Neighborhood based in the $x_{mix}$ vs. $U$ plane using low-mass stars. We can examine the other properties of these structures by considering additional parameters not used to initially identify the streams, i.e. photometric metallicity ($[M/H]_{Photo}$) and vertical velocity (W). Figure \ref{fig:metal_W_UV_plane} shows the median $[M/H]_{Photo}$ (left panel) and $\sigma_W$ (right panel) in the $x_{mix}$ vs. $U$ plane, where ellipses for each stream at the $2\sigma$ level are shown for reference. In both panels, significant sub-structure is observed that matches the location of the identified kinematic groups.

For the median $[M/H]_{Photo}$ plot, it is clear that median metallicity inside some of the streams is higher or lower than other stars with similar angular momenta (i.e. $x_{mix}$). This is especially true for the Hyades stream (red ellipse). This effect has been observed by \citet{bovy2010}, where it was argued that streams with metallicity distributions that greatly differ from stars with similar guiding radii must have formed due to interactions with dynamical resonances, which caused these stars to have different birth radii, and thus chemical composition, than the other stars. \citet{bovy2010} concluded that only the Hyades steam exhibited evidence of such an origin based on metallicity, a conclusion that is qualitatively supported by Figure \ref{fig:metal_W_UV_plane}, but possible chemical differences are also observed for other streams.

Similar to the median $[M/H]_{Photo}$, the standard deviation in the vertical velocity ($\sigma_W$) also displays significant sub-structure, where the variations are associated with the location of some of the groups. For stars in many of the streams, the vertical velocity scatter is found to be much lower than for stars with similar guiding radii. A smaller scatter in the vertical velocity of a group of a sub-population can be attributed to the stars being young, an effect that has been observed in many studies, most notably in \citet{nordstrom2004}.

\subsection{Chemodynamical Age Calibration}\label{sec:age_cal}

As reported in Section \ref{sec:var_x_U}, variations in both the median metallicity and vertical velocity dispersion are observed across the $x_{mix}$ vs. $U$ plane, where many of these variations are correlated with the location of defined stellar streams. Qualitatively, some of these variations, most notably differences in the vertical velocity dispersion, can be associated with age. To quantify the relationship between the chemodynamics and age of a population of stars, we use data from stars of higher masses from GALAH DR3 for reference. The ages of the GALAH stars come from one of the value added catalogs that derived isochronal ages using the Bayesian Stellar Parameter Estimation code BSTEP \citep{sharma2018}. For this sample, we only consider ages with errors $<20\%$, stars where the flags flag\_sp$= 0$ and flag\_fe\_h$ = 0$, and stars within 500 pc.

To examine changes in metallicity and vertical velocity for mono-age populations, we examine plots of vertical velocity, $W$, vs. metallicity, $[Fe/H]$.

\begin{figure*}
	\includegraphics[width=0.49\textwidth]{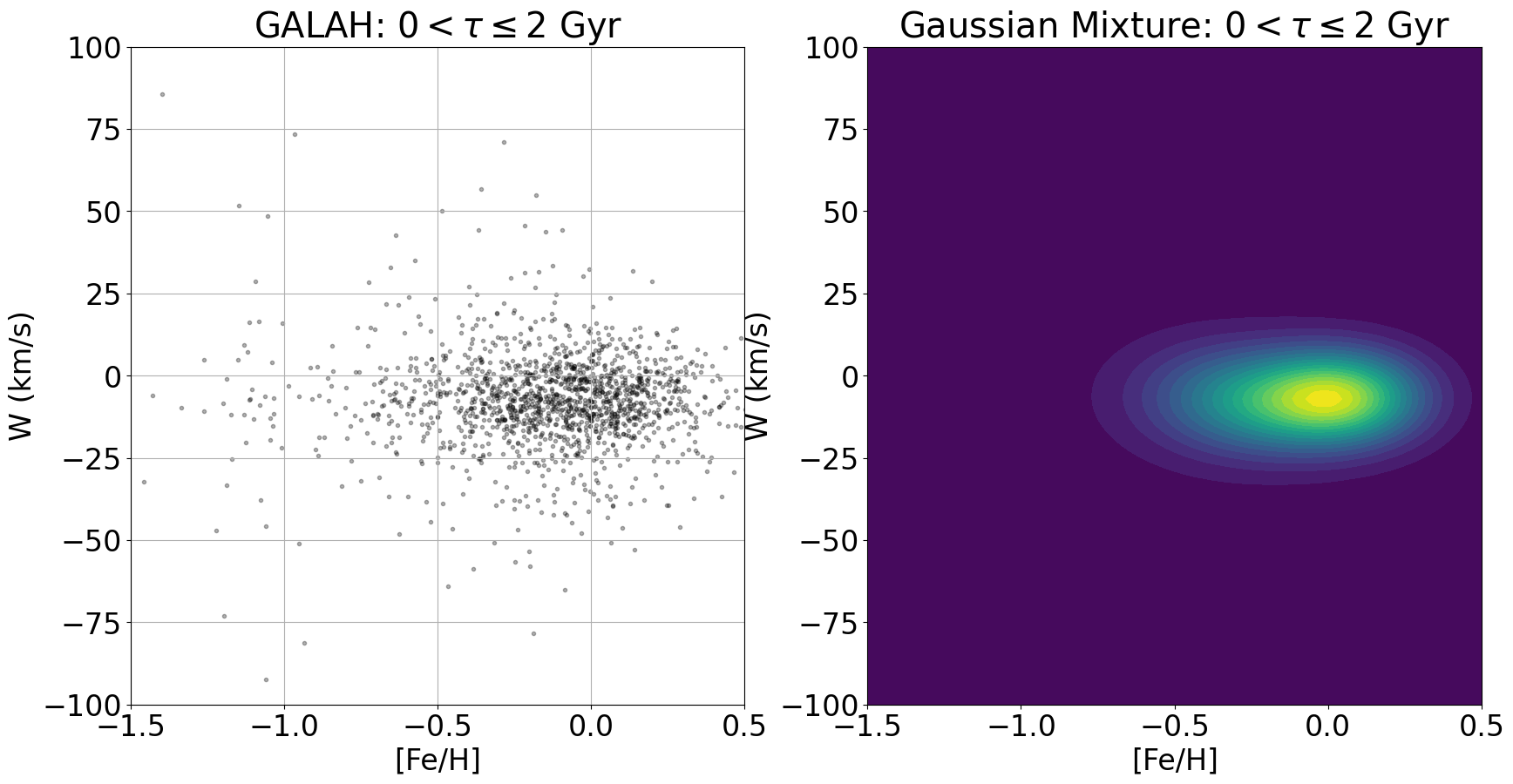}
	\includegraphics[width=0.49\textwidth]{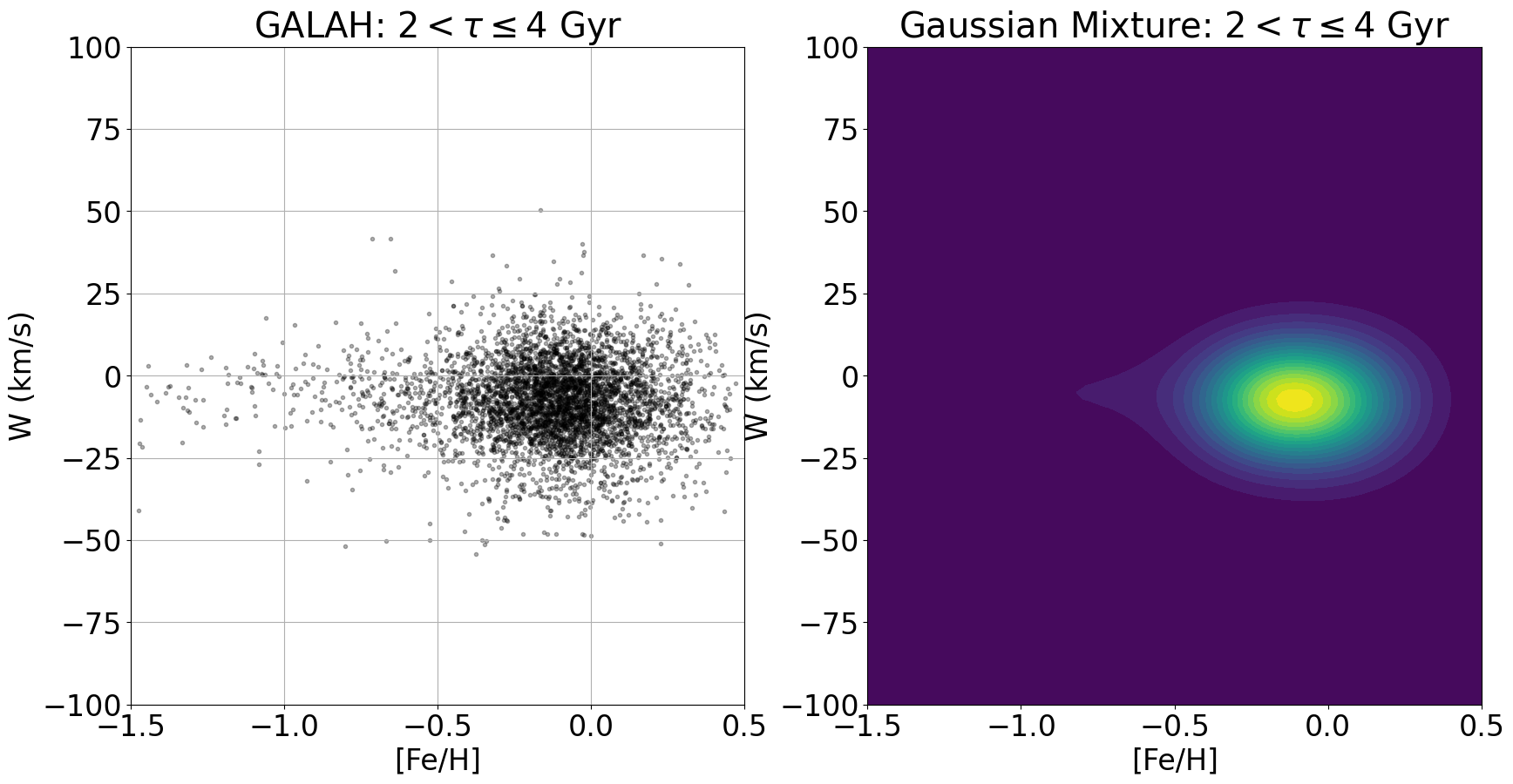}
	\includegraphics[width=0.49\textwidth]{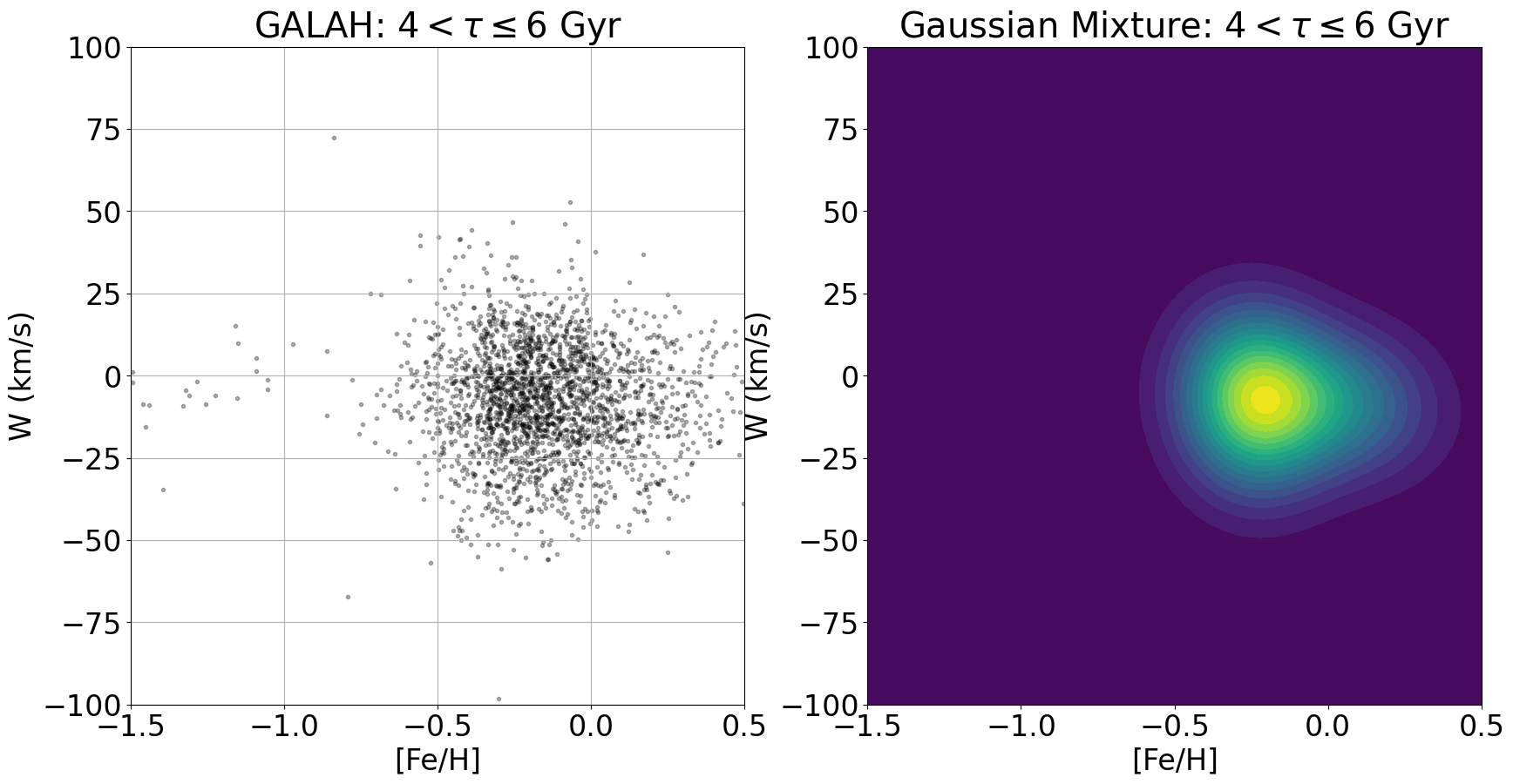}
	\includegraphics[width=0.49\textwidth]{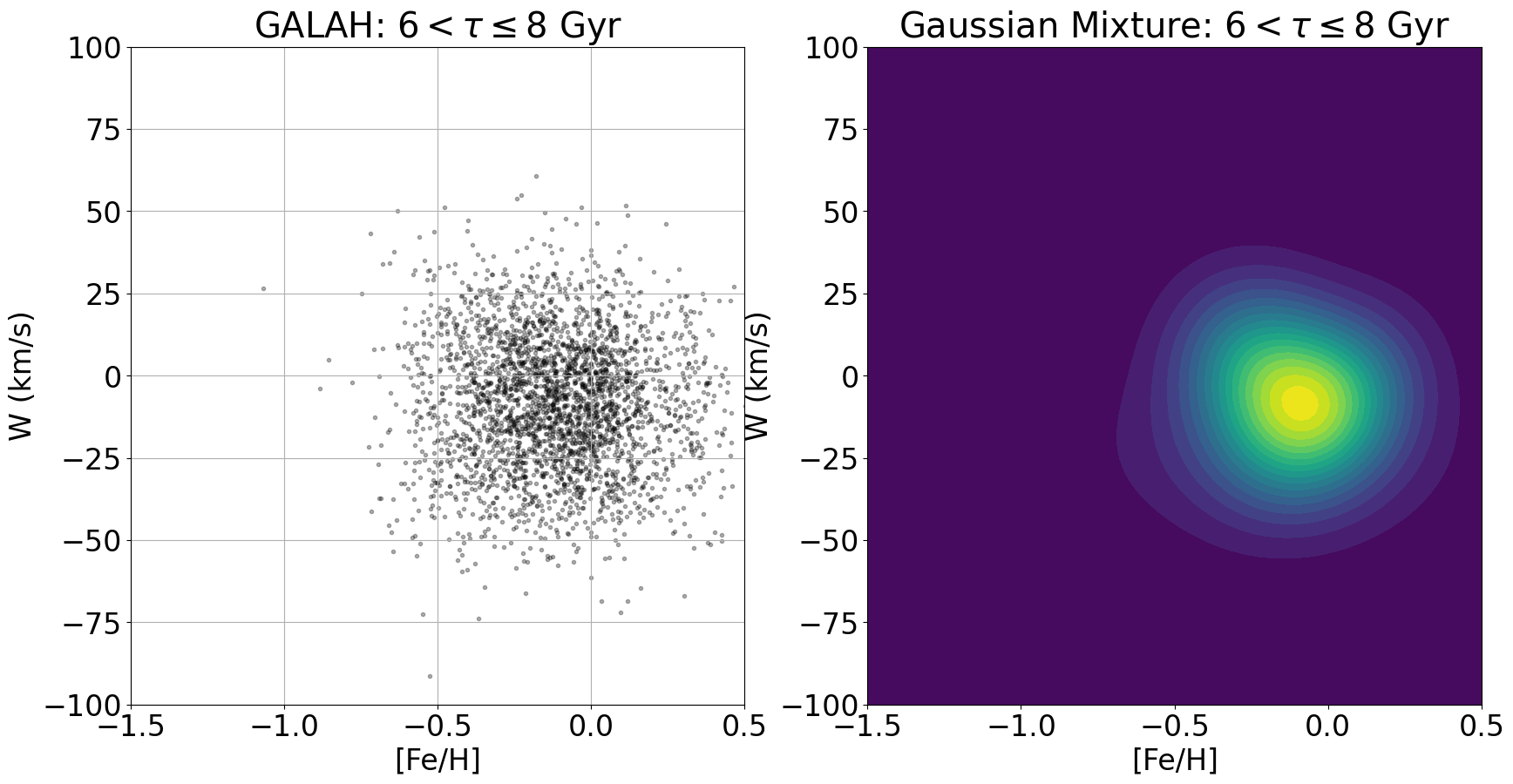}
	\includegraphics[width=0.49\textwidth]{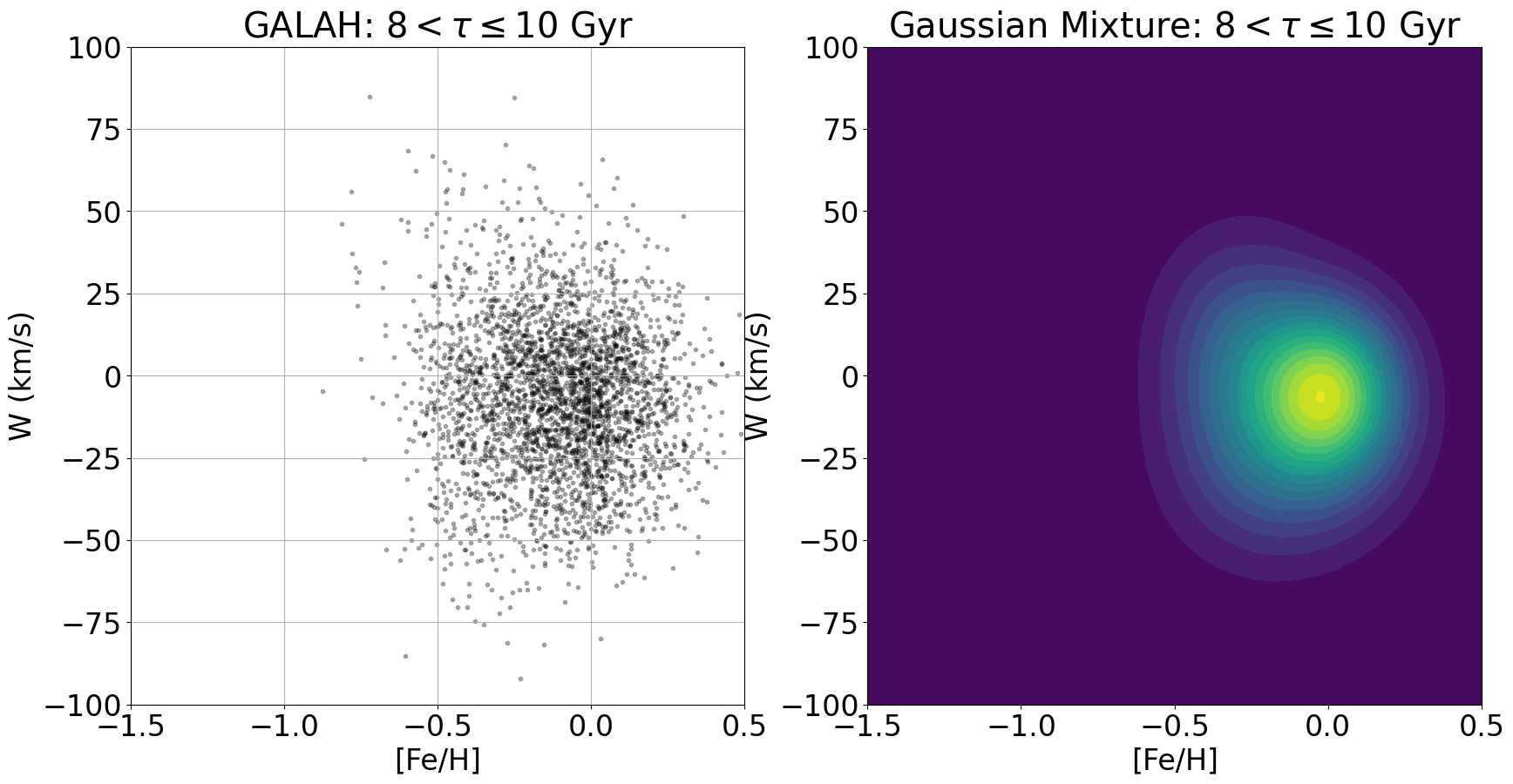}
	\includegraphics[width=0.49\textwidth]{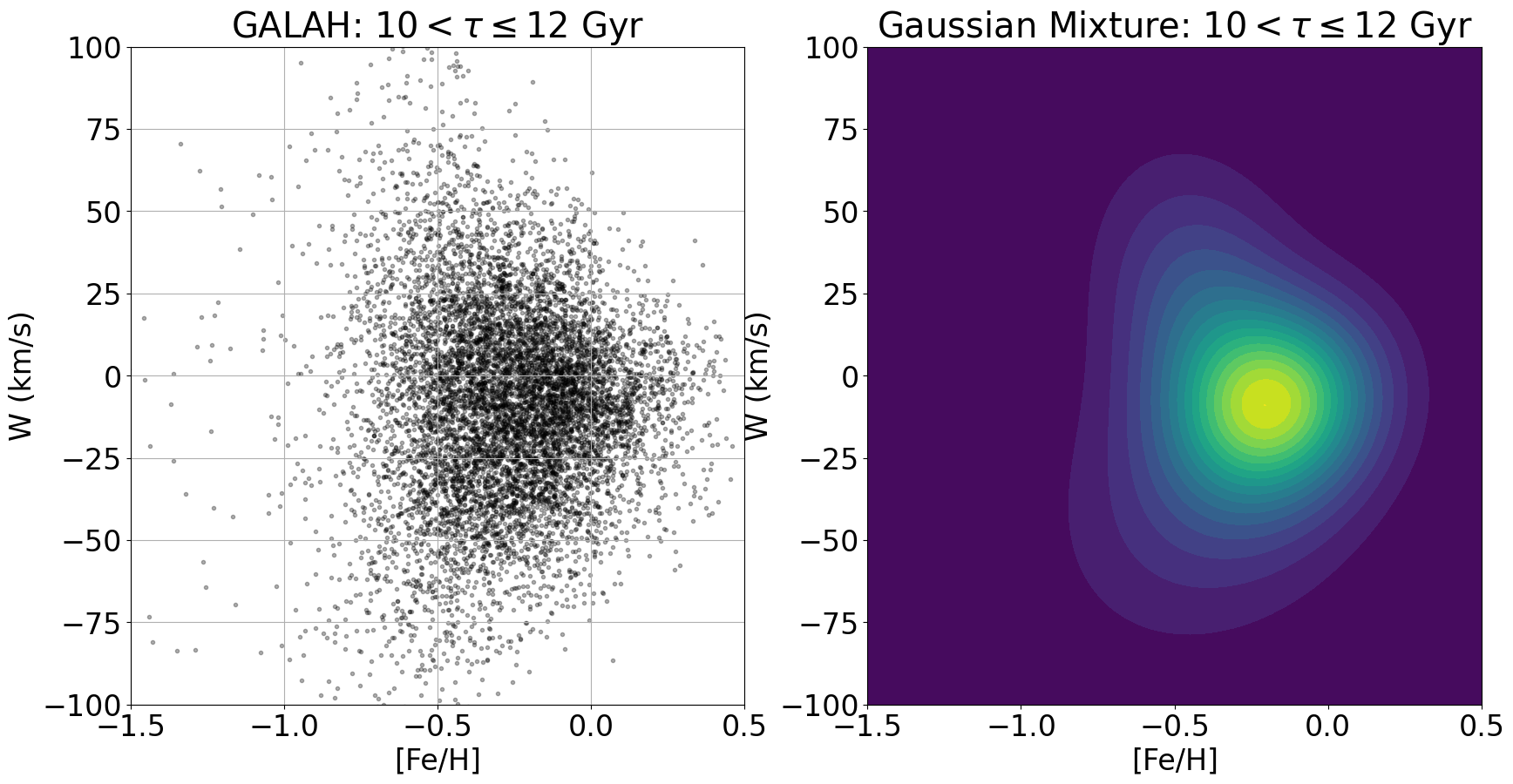}
	\includegraphics[width=0.49\textwidth]{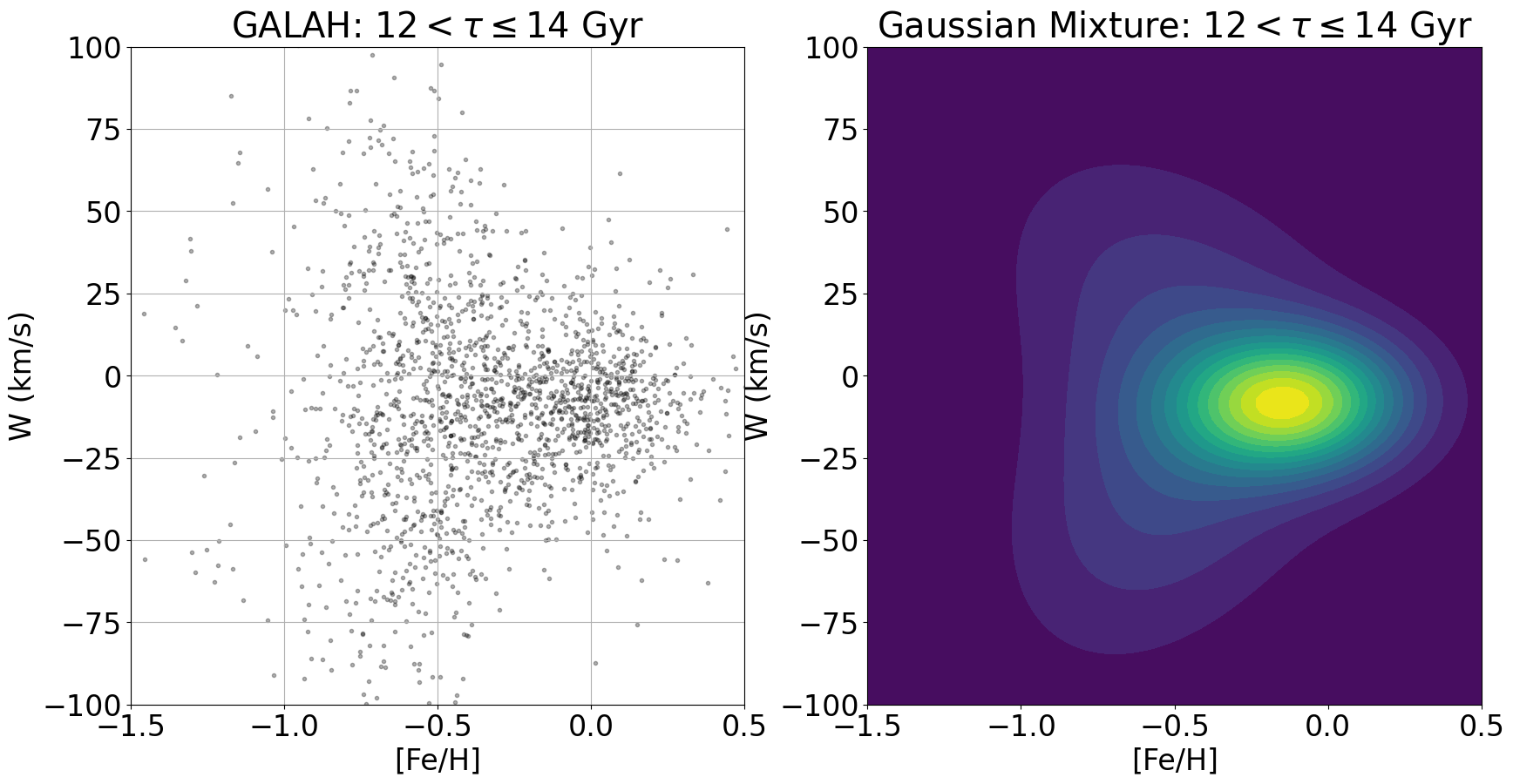}
	\caption{Plots of $W$ vs. $[Fe/H]$ for the stars in GALAH with isochronal ages in bins of 2 Gyr. The plots with the black data points show the observed distribution of stars for each mono-age population and the contour plots show the modeled distribution using a three-component Gaussian Mixture model. The distributions notably show a significant increase in W scatter with age along with a metallicity shift.}
	\label{fig:galah_mono_age}
\end{figure*}

To model the probability distribution in the $W$ vs. $[Fe/H]$ plane for each of these mono-age populations, we use a three component Gaussian Mixture model, as implemented in \textit{sklearn} \citep{sklearn}. Figure \ref{fig:galah_mono_age} shows the observed (data points) and modeled (contour plots) distributions for stars in bins of 2 Gyr. As the age of the population increases, we see that the dispersion in $W$ velocity increases and the number of metal-poor stars also increases, similar to what has been observed in other studies \citep[e.g.][]{nordstrom2004}. In this two-dimensional view however, we see that this increase in $W$ disperison is metallicity dependent: the dispersion increases significantly for metal-poor stars as the population becomes older. This trend has been observed in previous studies that also examined the relation between metallicity and velocity dispersion for stars in spectroscopic surveys \citep{Sharma2014, bland2019}.

Using these Gaussian Mixture models, we can determine the most probable underlying age distribution of a population of lower-mass stars based on its observed $W$ vs. $[Fe/H]$ distribution, assuming that stars of high and low masses in a given population share similar chemical composition and kinematics. To do this, we assume that the overall observed age distribution, $F$, of a subset of low-mass stars can be represented by the sum of Gaussian Mixture models for each mono-age population:
\begin{equation}\label{eq:mod_eq}
	F(\vec{x}) = \sum_t^n  f_t \times \left( \sum_{i=1}^3 w_{i,t} N(\vec{x}, \mu_{i,t}, \Sigma_{i,t}) \right)
\end{equation}
where $f_t$ is the fraction of the total population from the observed sample in the mono-age population $t$, $w_{i,t}$ is the weight of the \textit{i}th distribution in the Gaussian Mixture model for the mono-age population $t$, and $N$ is a normal distribution for the \textit{i}th competent of the Gaussian Mixture model with some mean, $\mu_{i,t}$, and covariance, $\Sigma_{i,t}$. Using the results of our Gaussian Mixture models, we then only need to find the optimal $f_t$ values for a given distribution to provide the most probable underlying age distribution. This result will only provide a rough estimate of the age distribution, as we have parsed the GALAH sample in bins of 2 Gyr.

To find the most probable fractions for each age bin, we bootstrap the observed distribution of $W$ vs. $[Fe/H]$ for $N=1000$ iterations to get the uncertainties on the distribution, and then find the optimal model distribution from eq. \ref{eq:mod_eq} using a Markov chain Monte Carlo (MCMC) method, as implemented in the Python package \textit{emcee} \citep{emcee}. For the MCMC optimization, we use 100 walkers and let it run for 10,000 steps. We minimize the log likelihood during the run, such that:
\begin{equation}
	ln(p(y|x)) = -\frac{1}{2} \sum_i \left[\frac{(y_i - F(\vec{x_i}))^2}{\sigma_i^2} +ln(\sigma_i^2) \right]
\end{equation}
We also enforce a prior such that $f_t>0$ and $\sum_t^n f_t =1$. At the end of the run, we estimate the integrated auto-correlation time ($t_{AC}$; number of steps for walkers to forget where they started), and discard $3 t_{AC}$ steps and thin the sample by $t_{AC}/2$ to then calculate the most likely age distribution, along with the errors on the distribution based on the 16th and 84th percentiles.

\begin{figure*}
	\centering
	\begin{subfigure}[b]{0.49\textwidth}
		\includegraphics[width=\textwidth]{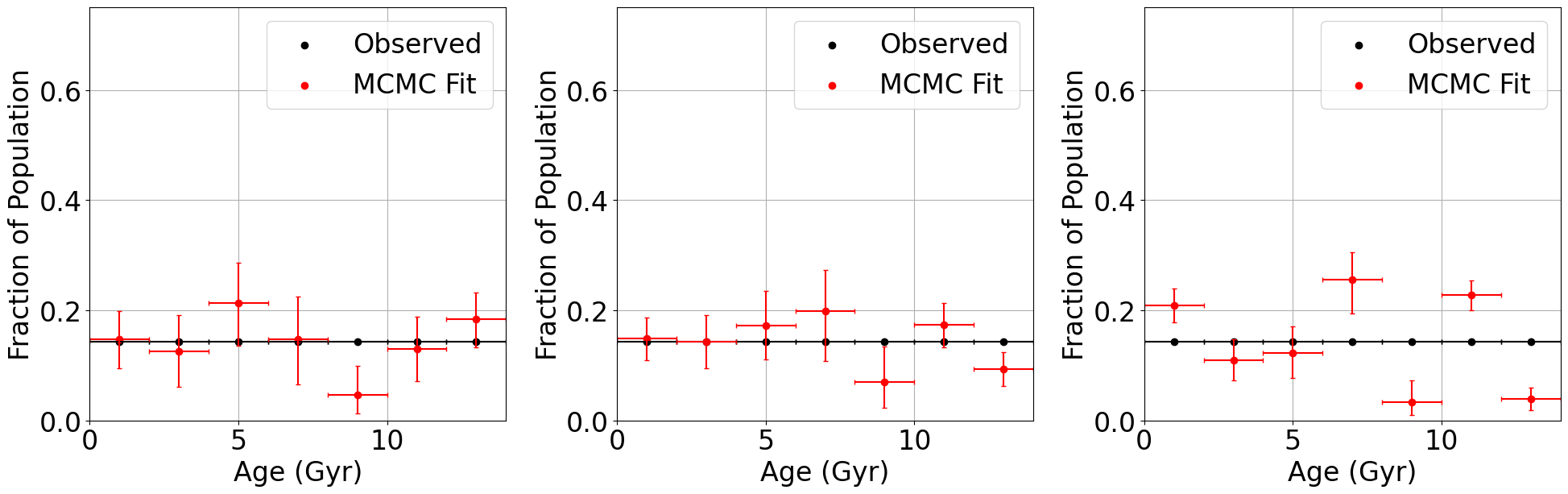}
		\caption{Single-Peak 1}
		\label{fig:galah_one_peak_test_A}
	\end{subfigure}
    \quad
    \begin{subfigure}[b]{0.49\textwidth}
    	\includegraphics[width=\textwidth]{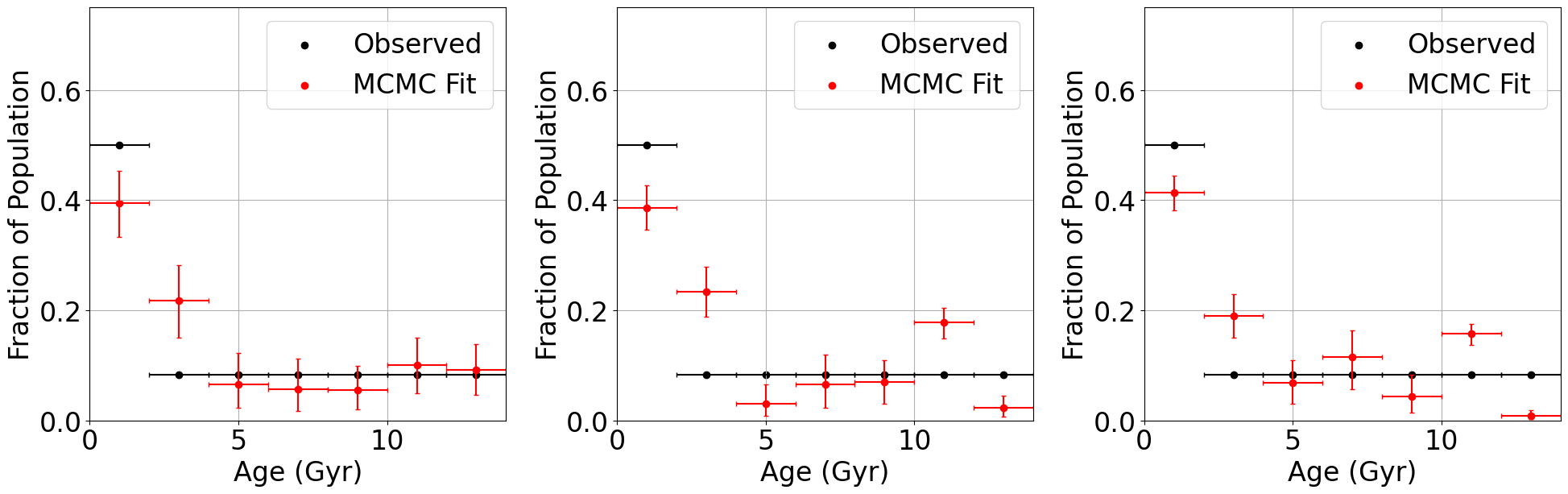}
    	\caption{Single-Peak 2}
    	\label{fig:galah_one_peak_test_B}
    \end{subfigure}
    \quad
    \begin{subfigure}[b]{0.49\textwidth}
    	\includegraphics[width=\textwidth]{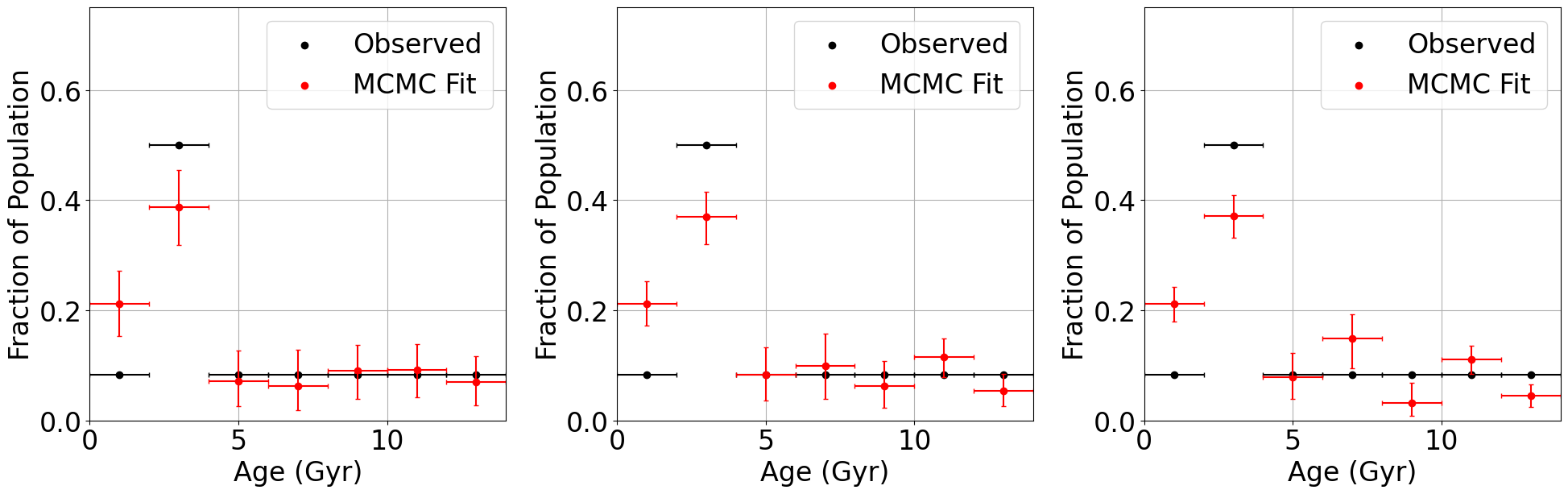}
    	\caption{Single-Peak 3}
    	\label{fig:galah_one_peak_test_C}
    \end{subfigure}
    \quad
    \begin{subfigure}[b]{0.49\textwidth}
    	\includegraphics[width=\textwidth]{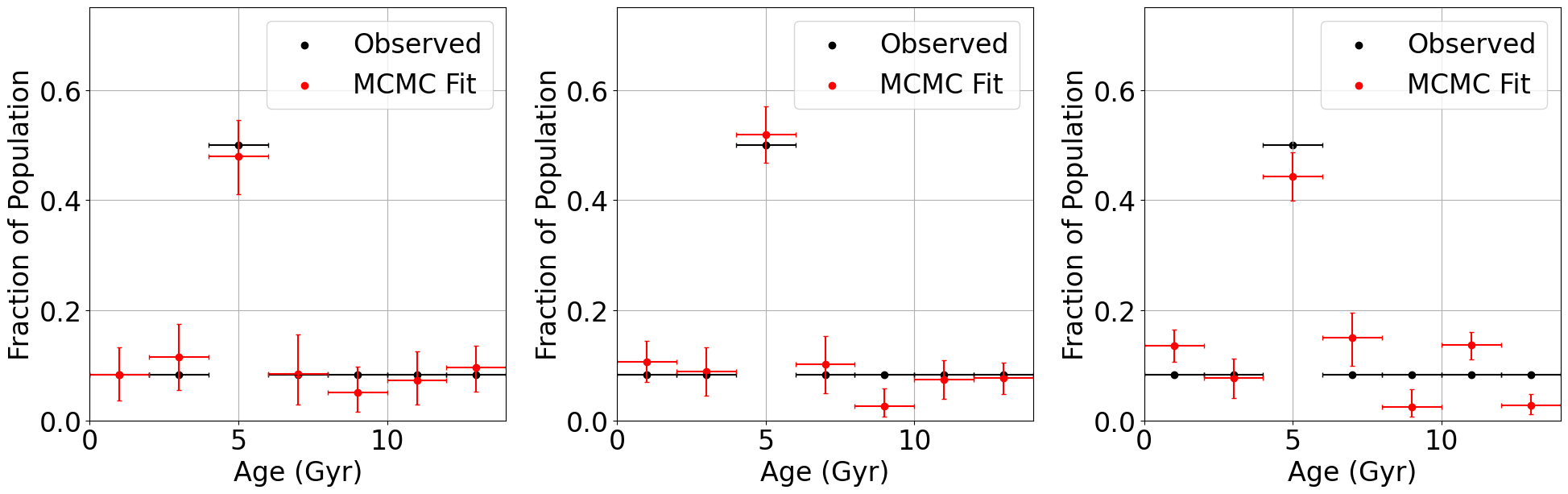}
    	\caption{Single-Peak 4}
    	\label{fig:galah_one_peak_test_D}
    \end{subfigure}
    \quad
    \begin{subfigure}[b]{0.49\textwidth}
    	\includegraphics[width=\textwidth]{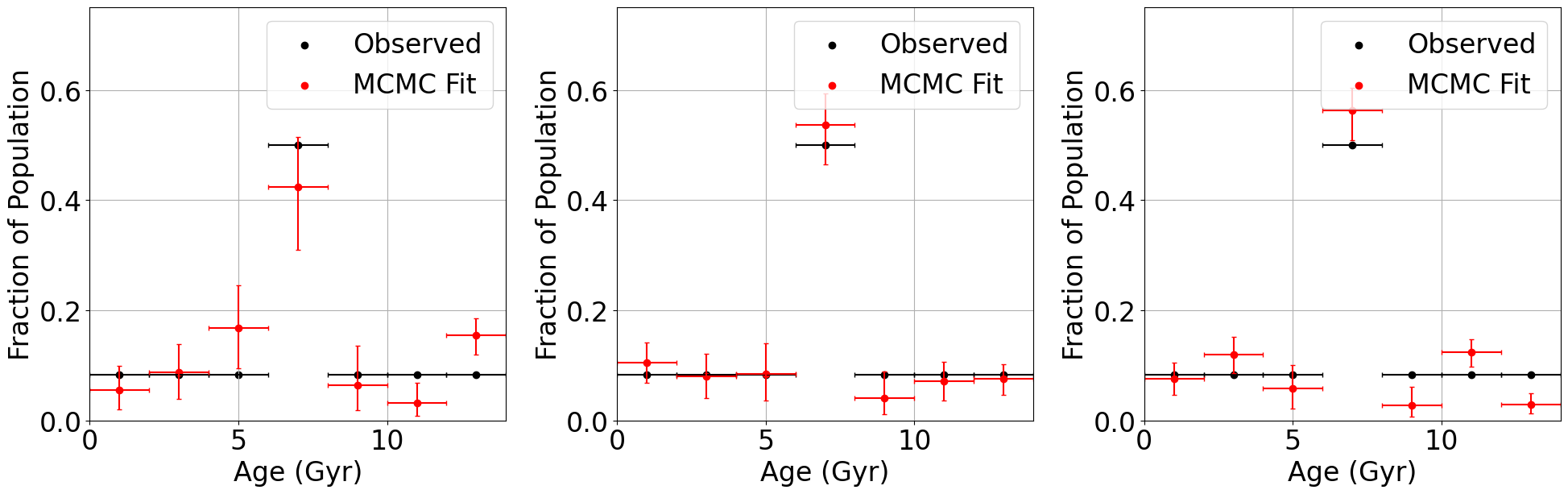}
    	\caption{Single-Peak 5}
    	\label{fig:galah_one_peak_test_E}
    \end{subfigure}
    \quad
    \begin{subfigure}[b]{0.49\textwidth}
    	\includegraphics[width=\textwidth]{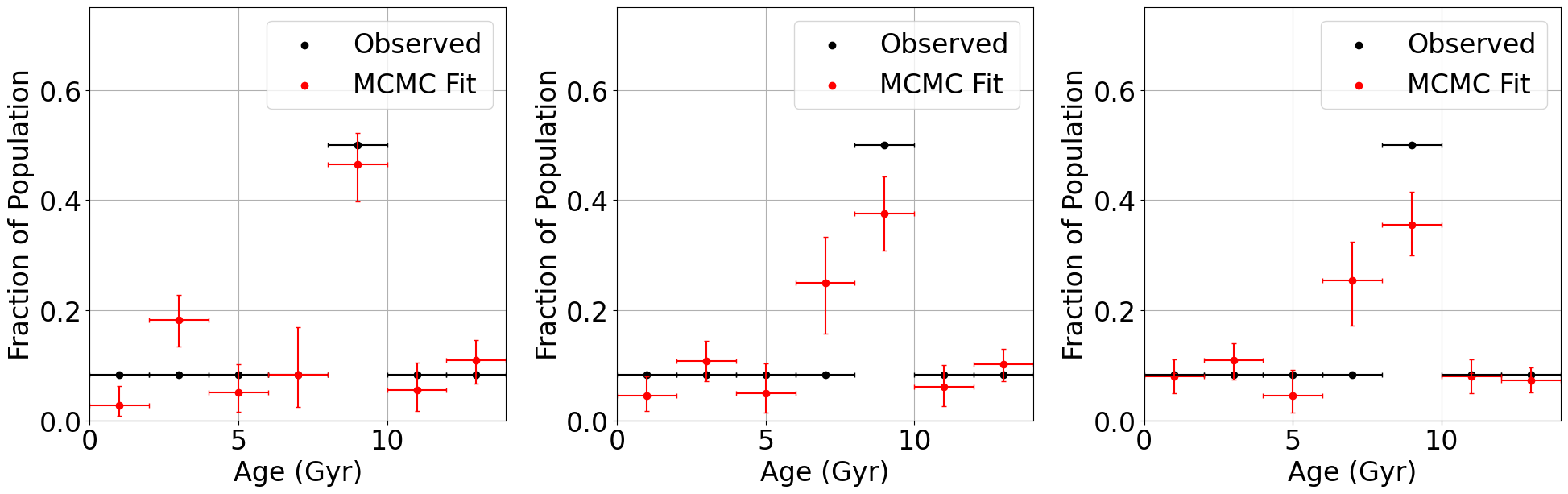}
    	\caption{Single-Peak 6}
    	\label{fig:galah_one_peak_test_F}
    \end{subfigure}
    \quad
    \begin{subfigure}[b]{0.49\textwidth}
    	\includegraphics[width=\textwidth]{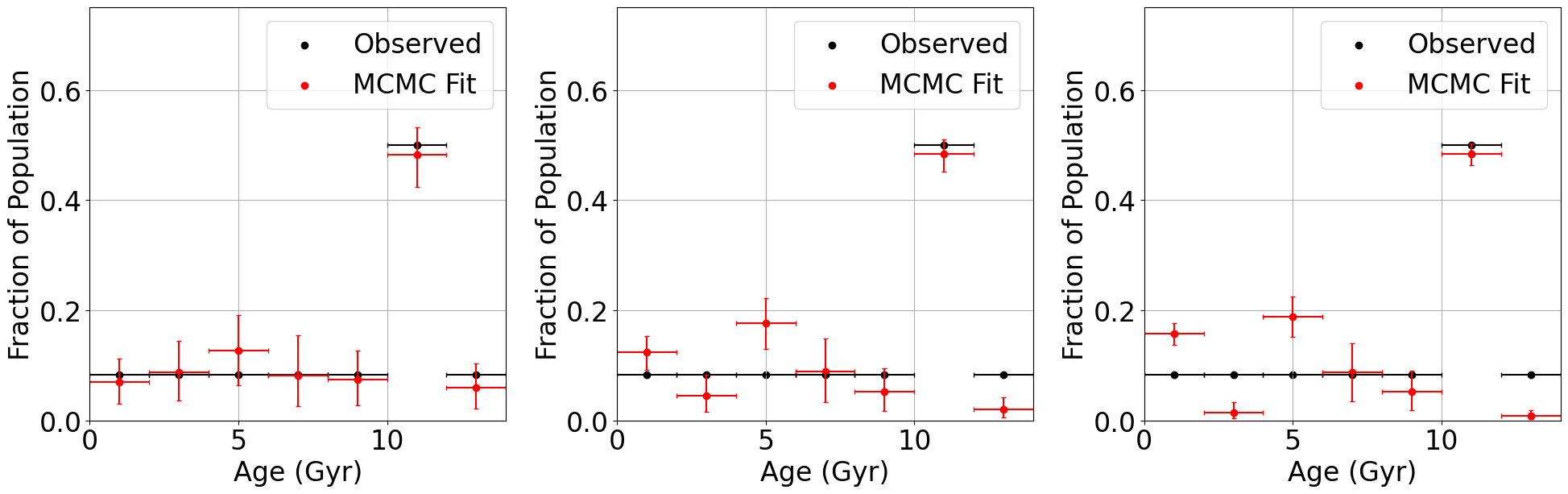}
    	\caption{Single-Peak 7}
    	\label{fig:galah_one_peak_test_G}
    \end{subfigure}
    \quad
    \begin{subfigure}[b]{0.49\textwidth}
    	\includegraphics[width=\textwidth]{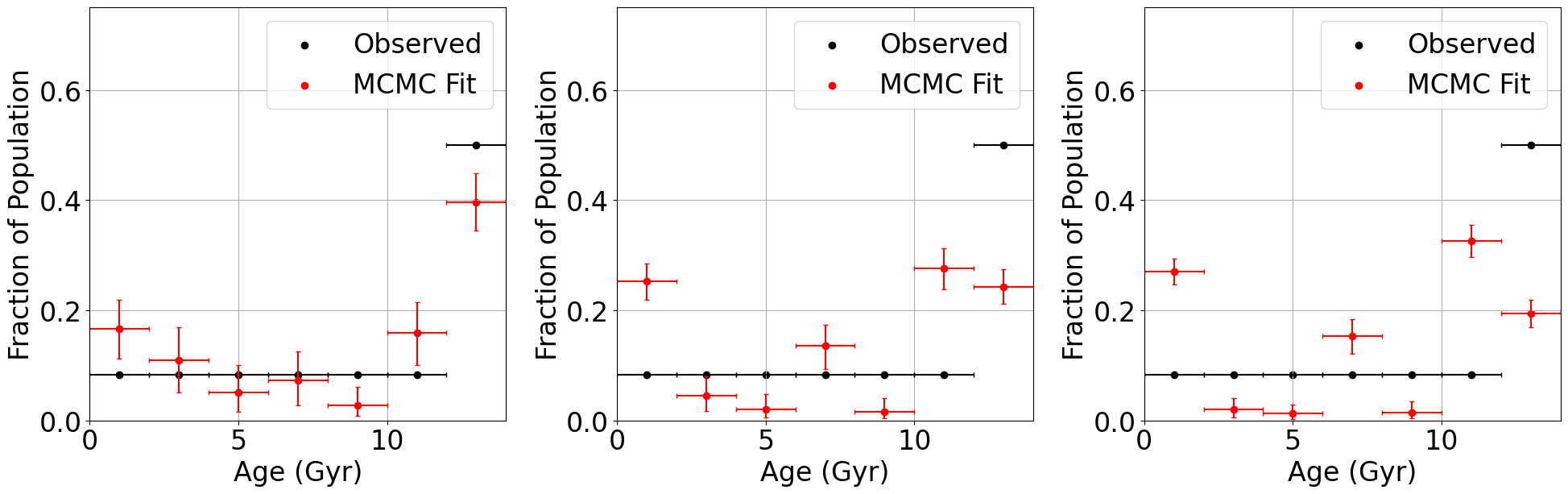}
    	\caption{Single-Peak 8}
    	\label{fig:galah_one_peak_test_H}
    \end{subfigure}
	\caption{Numerical attempts to recover the underlying age distribution from fits of the metallicity and kinematics distribution in $W$ vs. $[Fe/H]$, for groups of GALAH stars of known ages with a single "peak" in age distribution. For each attempt (in red), the MCMC optimization is performed on three different subsets that have the same underlying age distribution shown in black.}
	\label{fig:galah_one_peak_test}
\end{figure*}

\begin{figure*}
	\centering
	\begin{subfigure}[b]{0.49\textwidth}
		\includegraphics[width=\textwidth]{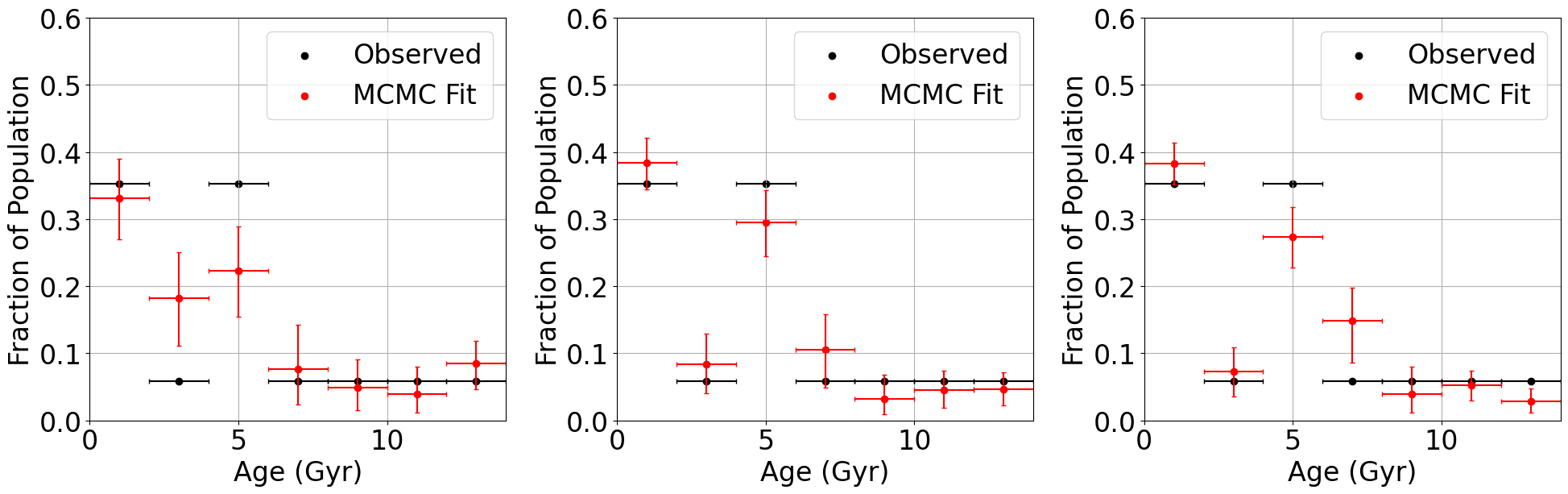}
		\caption{Dual-Peak 1}
		\label{fig:galah_two_peak_test_A}
	\end{subfigure}
	\quad
	\begin{subfigure}[b]{0.49\textwidth}
		\includegraphics[width=\textwidth]{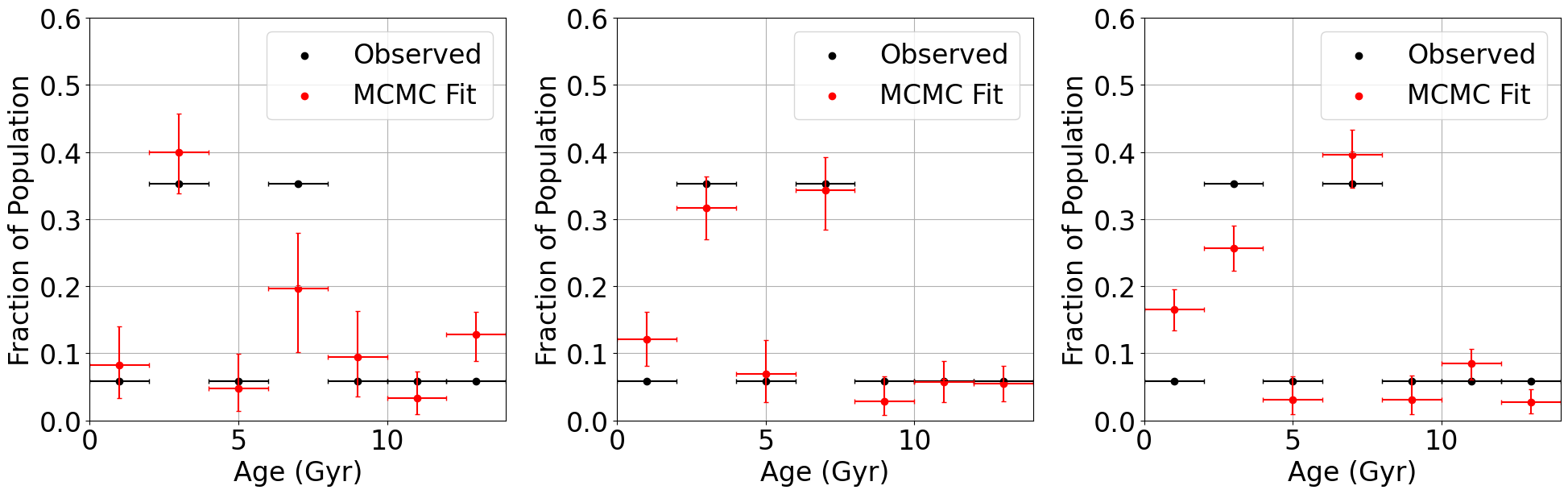}
		\caption{Dual-Peak 2}
		\label{fig:galah_two_peak_test_B}
	\end{subfigure}
	\quad
	\begin{subfigure}[b]{0.49\textwidth}
		\includegraphics[width=\textwidth]{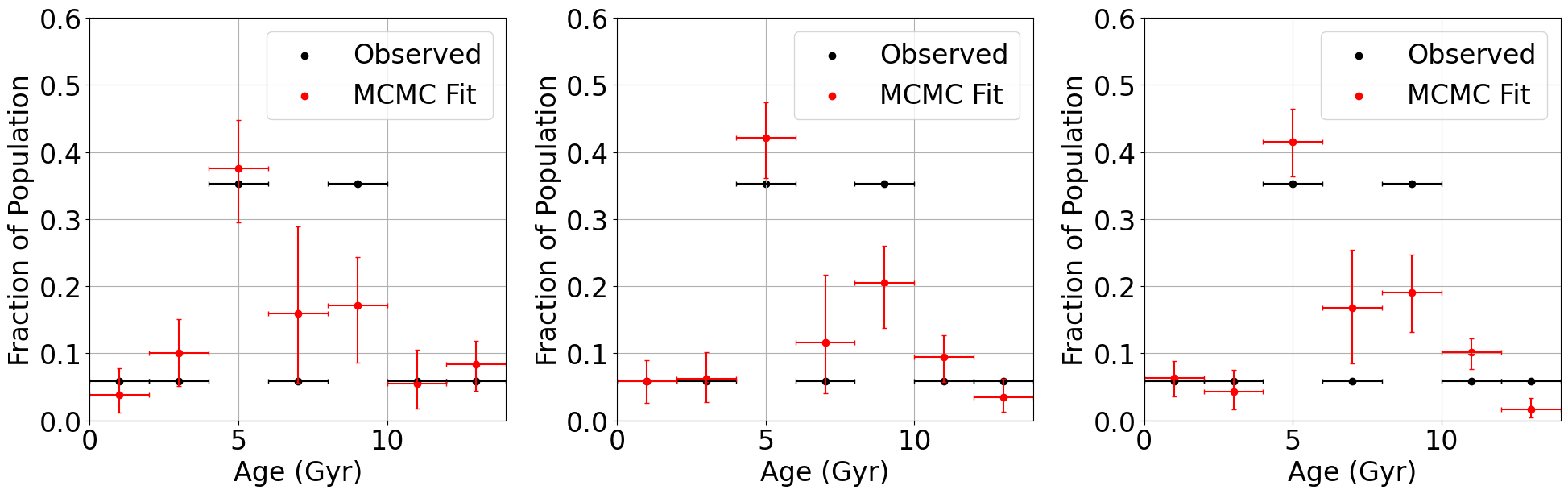}
		\caption{Dual-Peak 3}
		\label{fig:galah_two_peak_test_C}
	\end{subfigure}
	\quad
	\begin{subfigure}[b]{0.49\textwidth}
		\includegraphics[width=\textwidth]{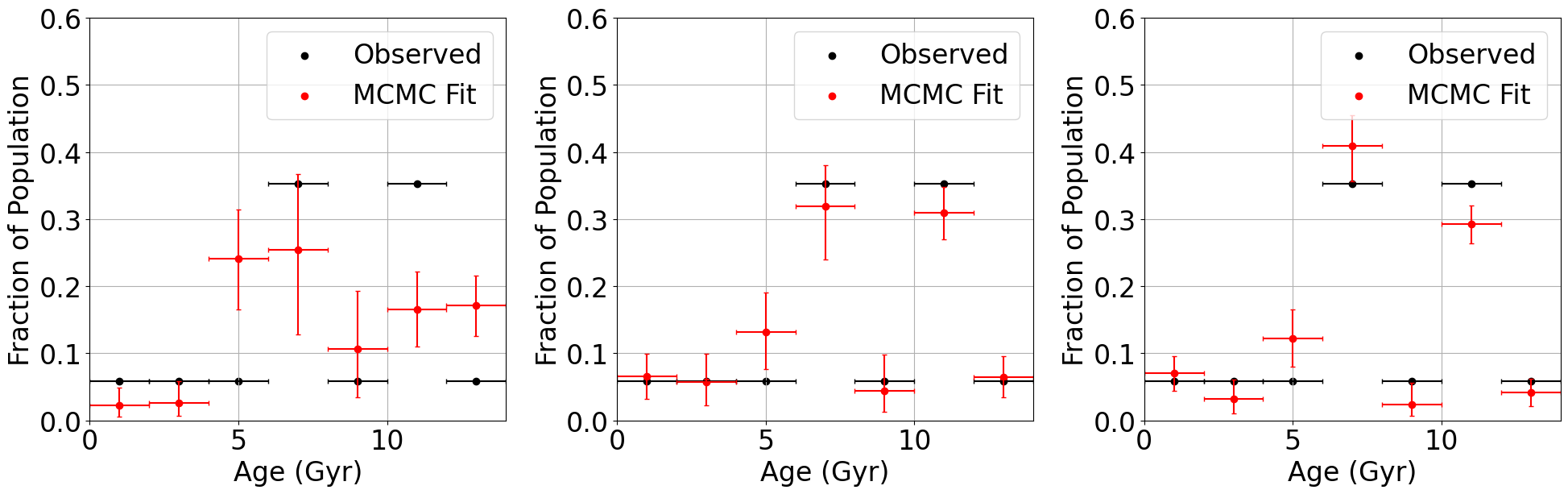}
		\caption{Dual-Peak 4}
		\label{fig:galah_two_peak_test_D}
	\end{subfigure}
	\quad
	\begin{subfigure}[b]{0.49\textwidth}
		\includegraphics[width=\textwidth]{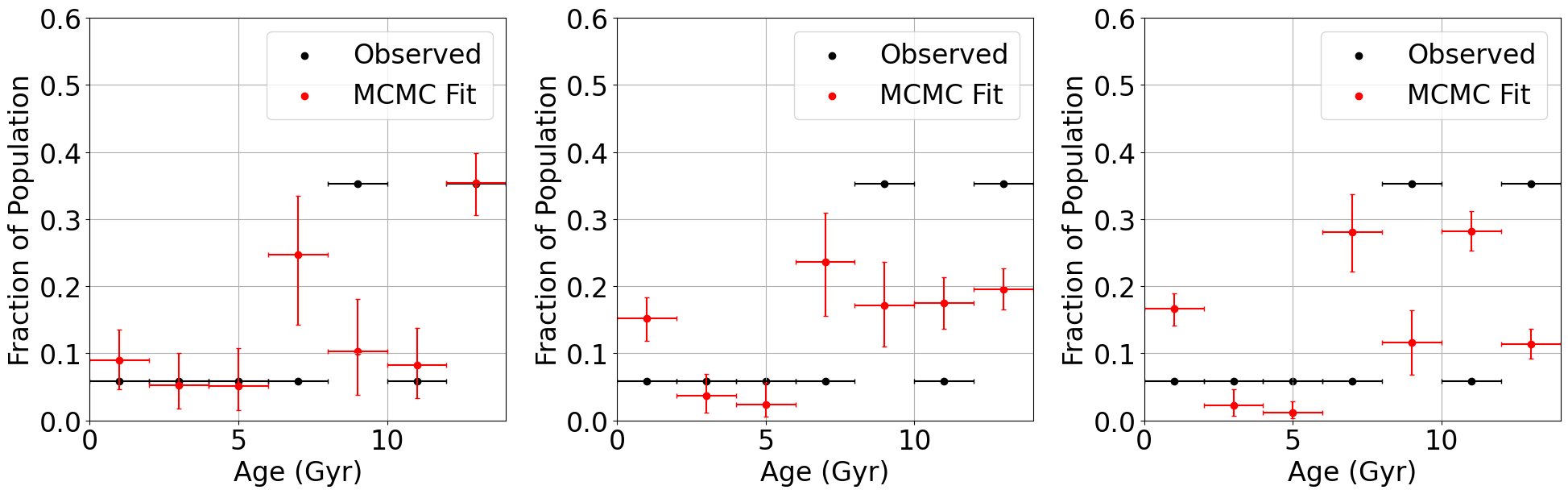}
		\caption{Dual-Peak 5}
		\label{fig:galah_two_peak_test_E}
	\end{subfigure}
	\caption{Numerical attempts to recover the underlying age distribution from fits of the metallicity and kinematics distribution in $W$ vs. $[Fe/H]$, for groups of GALAH stars of known ages with two "peaks" in their age distribution. For each attempt (in red), the MCMC optimization is performed on three different subsets that have the same underlying age distribution shown in black.}
	\label{fig:galah_two_peak_test}
\end{figure*}

To examine the effectiveness of these fits, we use test sets from GALAH DR3 where various shapes of the age distribution were generated. For all the tests, the sample size is set to $N\approx 2,500$ and the data for each age bin is otherwise randomly chosen from the GALAH sample. Figures \ref{fig:galah_one_peak_test} and \ref{fig:galah_two_peak_test} show the results of this MCMC optimization applied to these random subsets of stars of known underlying age distributions. In each case we compare the numerically inferred distribution (red points) to the true underlying distribution of the subset (black points). For each test distribution, we make separate attempts on three different subsets that have the same underlying age distribution to demonstrate what, if any, variations in the solution are present.

Focusing on the test distributions with a single peak (Figure \ref{fig:galah_one_peak_test}), we find that in all attempts, with the exception of Single-Peak 8, we can successfully recover the location of the single peak. Additionally, we find that in most attempts, with the exception of Single-Peak 3 and Single-Peak 8, we were able to successfully recover the relative fraction of stars in most age bins within $\sim 2\sigma$. With this we conclude that our method is able to recover the location in peaks of an age distribution and that we should be able to trust any differences when comparing age distributions if they differ by at least $2\sigma$. The exception to this seems to be for stars of the oldest population ($>10$ Gyr), which in any case we expect to make up only a small number of the actual distribution of stars in the Solar Neighborhood.

When considering the test distributions with two peaks (Figure \ref{fig:galah_two_peak_test}), we again find that our method is able to reliably determine both the location of the peaks and the relative fraction of stars in most age bins to within $\sim 2\sigma$. The exception to this is for the Dual-Peak 5 case, and in some runs of the Dual-Peak 3 and 4 cases; just as in the Single-Peak 8 case, the distribution for Dual-Peak 5 suggests possible issues with our method when determining age distributions for the oldest stars, which again make up a small number in our survey. Besides this, the issues with the Dual-Peak 3 and 4 cases seem to be mostly in the older of the two peaks in each case, where the mthod cannot consistentlyt determine the relative fraction of stars are the $2\sigma$ level. This is again reserved mostly for the older age bins, while the rest of the bins still provide an accurate estimate (within $\sim 2\sigma$) for most of the age bins, which is consistent with our previous conclusions on the accuracy of this method.

\subsubsection{Inferred Birth Radii}\label{sec:birth_radii}

We know that disk stars in the Solar Neighborhood may have originated from Galactic radii much smaller or larger than their current location, in a process dubbed radial migration by \citet{sellwood2002}. This process can either be a ``cold" or ``hot" process, where in a cold process the star's orbit changes size (and angular momentum) without change in its vertical extent (and eccentricity), while a hot process results in changes in an orbit's vertical extent (and eccentricity). \citet{frankel2020} found that the cold process dominates in the Solar Neighborhood based on the kinematics, metallicity and age of APOGEE red clump stars. 

Using the assumptions from \citet{frankel2020}, in combination with the observed $[Fe/H]$ and modeled age distributions, we can infer the most probable birth radii distribution for a population of stars. To see the full explanation of the models used to describe the orbital evolution of stars, see \citet{frankel2020}. Here we provide a summary of the model relevant to the goal of inferring birth radii. 

In this model from \citet{frankel2020}, it is assumed that (1) stars are born with a tight relation between their metallicity and Galactic orbital angular momentum at any given time \citep{Krumholz2018, ness2019}, (2) secular processes dominate during the evolution of the Milky Way's thin disk, (3) the current Galactic potential can be approximated from the \textit{MWPotential2014} from \textit{galpy} \citep{galpy} and (4) action coordinates for an orbit can be estimated with a St\"{a}ckel approximation as implemented in \textit{galpy}.

To estimate the brith radius of a star, we first determine the initial angular momentum of a star, $L_{Z0}$, by inverting eq. 7 from \citet{frankel2020}. This results in:
\begin{widetext}
\begin{equation}\label{eq:most_prob_LZ}
	L_{Z0} = \begin{cases}
		\frac{235 \ km \ s^{-1}}{\nabla \text{inner}} \left[ [Fe/H] -  [Fe/H]_{max} f(\tau) \right], & \text{if $[Fe/H] > [Fe/H]_{max} f(\tau) + 3 \ \nabla \text{inner}$}\\
			\frac{235 \ km \ s^{-1}}{\nabla [Fe/H]} \left[ [Fe/H] -  [Fe/H]_{max} f(\tau) - 3 \left(\nabla \text{inner} - \nabla [Fe/H] \right) \right], & \text{if $[Fe/H] \leq [Fe/H]_{max} f(\tau) + 3 \ \nabla \text{inner}$}
	\end{cases}
\end{equation}
\end{widetext}
where:
\begin{equation}
	f(\tau) = \left( 1 - \frac{\tau}{12 \ Gyr} \right)^{\gamma_{[Fe/H]}} 
\end{equation}
In the above, the ``inner" metallicity gradient, $\nabla \text{inner}$, is fixed at $-0.03 / dex/kpc$, and the outer gradient ($\nabla [Fe/H]$), the maximum Galactic center metallicity ($[Fe/H]_{max}$) and the rate of change of metallicity with time ($\gamma_{[Fe/H]}$) are fitted parameters \citep[see Figure 4 in][]{frankel2020}. Finally, assuming a flat rotation curve with a circular velocity of $235 \ km \ s^{-1}$, the birth radius of a star can be estimated by:
\begin{equation}\label{eq:birth_R}
	R_{birth} = \frac{L_{Z0}}{235 \ km \ s^{-1}}
\end{equation}

In practice, using our method from Section \ref{sec:age_cal} we cannot know the individual age of a star, just the underlying distribution for a sample of stars. However, we can still use the discrete $[Fe/H]$ and $\tau$ distributions from a population to bootstrap the underlying birth radii distribution. To do this, we build a random sample of the same size as the population by pulling from discrete distributions with bin widths of $0.1$ dex for $[Fe/H]$, and $2$ Gyr for $\tau$, where the metallicity distribution is the average for the population bootstrapped for 1000 iterations and the age distribution is the most probable one found using the method above. Using this sample we can estimate birth radii using eqs. \ref{eq:most_prob_LZ}-\ref{eq:birth_R} and estimate the distribution of birth radii for the random sample. For each iteration random Gaussian errors are also added to the discrete $[Fe/H]$ and $\tau$ distributions based on variance from either the bootstrapping, in the case of the metallicity, or the MCMC, in the case of the age. This is done for 1000 iterations to bootstrap the most probable birth radii distribution for the population.

\section{Discussion}\label{sec:discuss}

\subsection{Chemodynamical Ages and Birth Radii Distributions}

In Section \ref{sec:age_cal} we demonstrated a method to estimate the most probable underlying age distribution of a group of stars based the group's $W$ vs. $[$M/H$]$ distribution. Additionally, we laid out a method in section \ref{sec:birth_radii} to use this age distribution, in combination with the metallicity, to infer a group of stars' birth radii distribution. Here we apply this method to groups of stars in the $x_{mix}$ vs. $U$ phase space. Specifically, we construct a grid consisting of rectangular regions of unequal areas such that the number stars in each region is approximately equal (specifically $N = 971 \pm 1$). To accomplish this, we use the KD-Tree algorithm as described in \citet{kdtree} and implemented in \textit{scipy} \citep{scipy}. In the 2D case, this algorithm divides the data at the median recursively for alternating axes. So, as the algorithm progresses, the previously created rectangles in the parameter space are recursively divided into equal number regions until some ``depth" which provides us approximately the number of stars per region desired. 

For each of these regions, we then estimate the age and birth radii distributions as described above.  When applying these methods to these kinematic groups, we are assuming that the underlying kinematic distribution and kinematic heating history for the more massive stars in GALAH used to test the method is the same as the lower mass K dwarfs used in this study. Also, we assume that the mean in the $W$ distribution does not vary from kinematic group to kinematic group. When finding the most probable age distribution for a group of stars, photometric metallicities have been corrected by:
\begin{equation}
	[M/H] = [M/H]_{photo} - 0.03
\end{equation}
where the additional correction of $-0.03$ has been added to account for the mean difference in metallicity between the GALAH stars and our photometric metallicities (Figure \ref{fig:compare_photo_metals}).

\begin{figure*}
	\centering
	\includegraphics[width=1\textwidth]{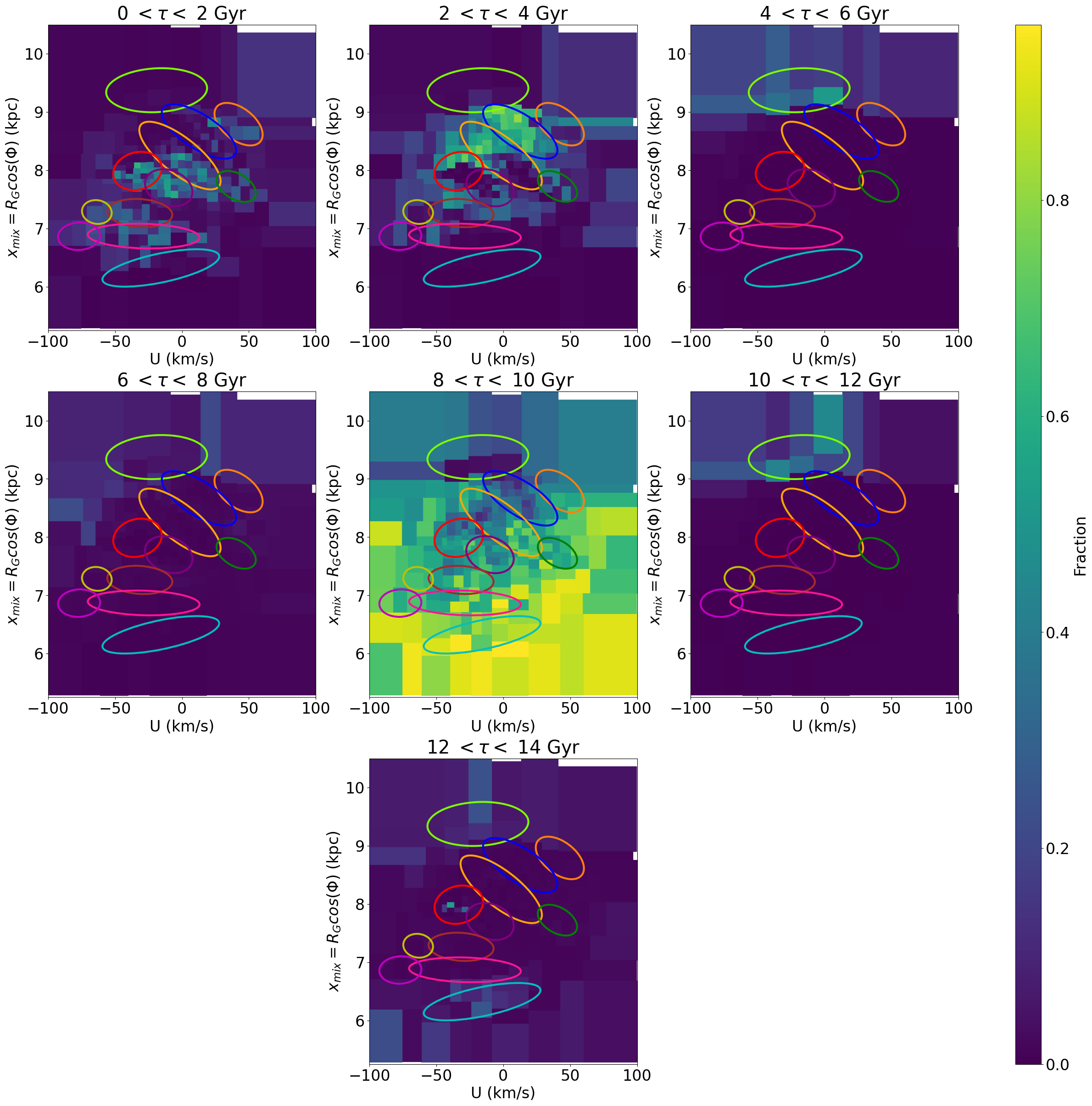}
	\caption{Fraction of stars from various age ranges in the $x_{mix}$ vs. $U$ phase space from the most probable age distributions determined using the methodology described in Section \ref{sec:age_cal}. The unequal area regions each contain an equal number of stars ($N = 971 \pm 1$) and are determined using the KD-Tree algorithm. The colored ellipses correspond to the kinematic groups identified in Section \ref{sec:kin_groups} and their colors correspond to the legend in Figure  \ref{fig:PCA_ellipses}. Stars from most kinematic groups appear to have high fractions of stars with ages $0<\tau<2$ and $2<\tau<4$ Gyr, except for the Hercules 3 group which appears to be dominated with stars of ages $8<\tau<10$ Gyr.}
	\label{fig:age_dist}
\end{figure*}

\begin{figure*}
	\centering
	\includegraphics[width=1\textwidth]{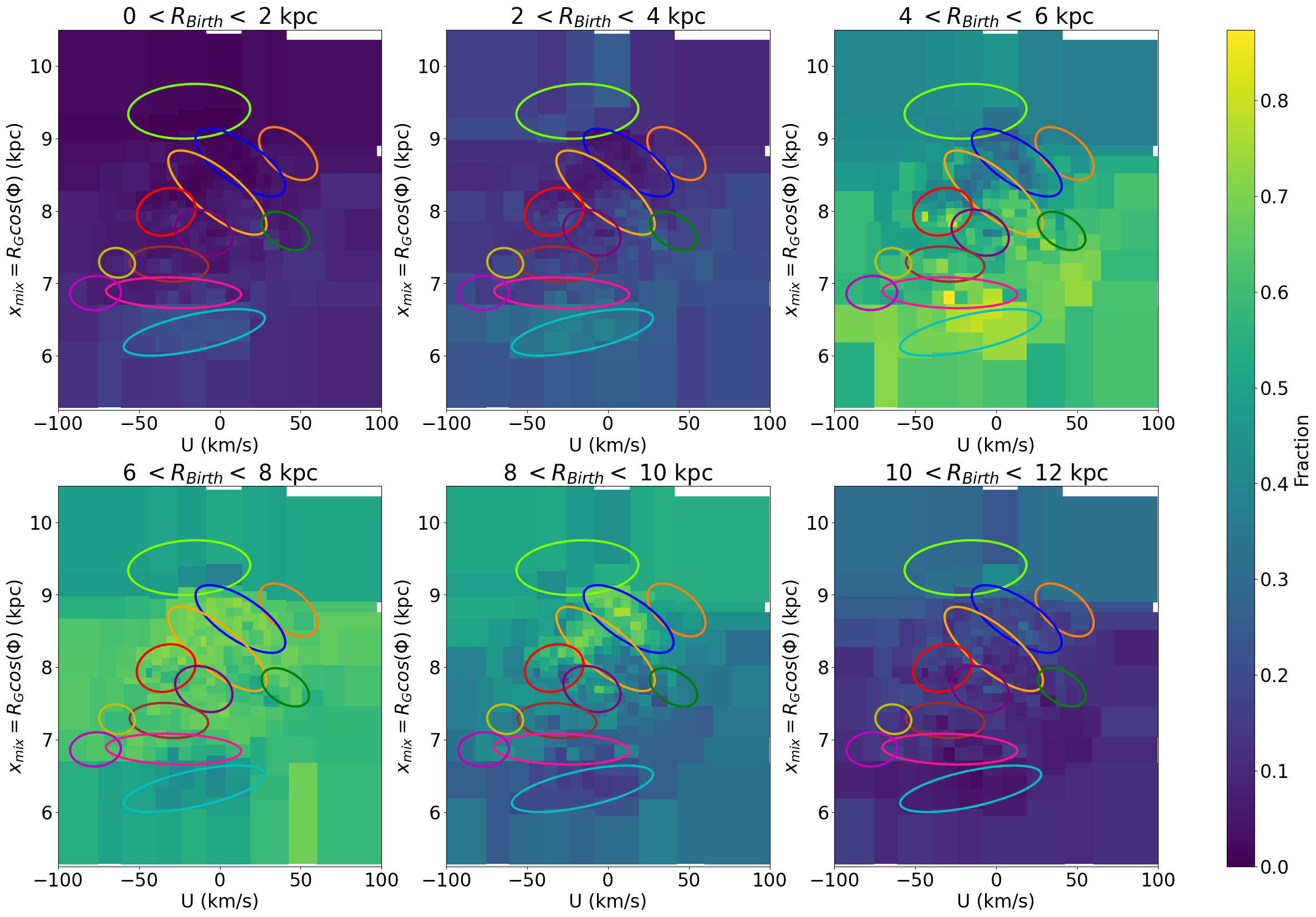}
	\caption{Fraction of stars with inferred birth radii $(R_{Birth})$ falling within various range bins in the $x_{mix}$ vs. $U$ phase space from the most probable inferred birth radii distributions determined from the methodology described in Section \ref{sec:birth_radii}. As a note, the inferred birth radii distributions are determined out to radii of 20 kpc, but only fractions out to 12 kpc are shown here as fractions at larger radii fall below $\sim 5\%$. The unequal area regions contain an equal number of stars ($N = 971 \pm 1$) and are determined using the KD-Tree algorithm. The colored ellipses correspond to the kinematic groups identified in Section \ref{sec:kin_groups} and the colors correspond to the legend in Figure  \ref{fig:PCA_ellipses}. Stars from within dense kinematic groups all appear to be local ($6<R_{Birth}<8$ and $8<R_{Birth}<10$ kpc), again with the exception of the Hercules groups whose stars appear to have originated closer to Galactic center ($0<R_{Birth}<6$ kpc for Hercules 3, and $4<R_{Birth}<8$ kpc for Hercules 1 and 2).}
	\label{fig:Rb_dist}
\end{figure*}

Figure \ref{fig:age_dist} and \ref{fig:Rb_dist} shows the fraction of stars in each bin in $x_{mix}$ vs. $U$ phase space that belong to populations of a certain age $\tau$ or inferred birth radii $R_{Birth}$, respectively. For the birth radii fractions in Figure \ref{fig:Rb_dist} we note that while distributions are determined out to radii of 20 kpc, only fractions out to 12 kpc are shown here, as fractions at larger radii fall below $\sim 5\%$. Additionally, when considering the errors in the fractions in these plots, the 50th (and 34th/86th) percentile of the uncertainties at the 95\% confidence interval are $0.047^{+0.069}_{-0.020}$ and $0.018^{+0.020}_{-0.010}$ for the most probable age and inferred birth radii fractions, respectively.

\subsubsection{Comparison With Past Observations}

When analyzing the distributions in Figures \ref{fig:age_dist} and \ref{fig:Rb_dist}, we first want to consider these results in the context of previous studies. From the probable age distributions in Figure \ref{fig:age_dist}, we find that generally most of the stars (in total) are found to be within age ranges of either 0$-$6 Gyr or 8$-$12 Gyr, at varying levels across the  $x_{mix}$ vs. $U$ plane. This result is consistent with the ``Two Infall Model" \citep{Chiappini1997}, which predicts two distinct periods of star formation for the disk. This is also observed in multiple studies of star formation in the Solar Neighborhood populations where it is found that maximum rates occurred at $\sim$2$-$5 Gyr ago and $\sim$9$-$10 Gyr ago \citep[e.g.][]{Cignoni2006,Snaith2015,mor2019,alzate2020}, which agrees quite well with the age ranges we find the most stars (i.e. $2-4$ Gyr and $8-10$ Gyr) in Figure \ref{fig:age_dist}.

\begin{figure}
	\centering
	\includegraphics[width=0.8\columnwidth]{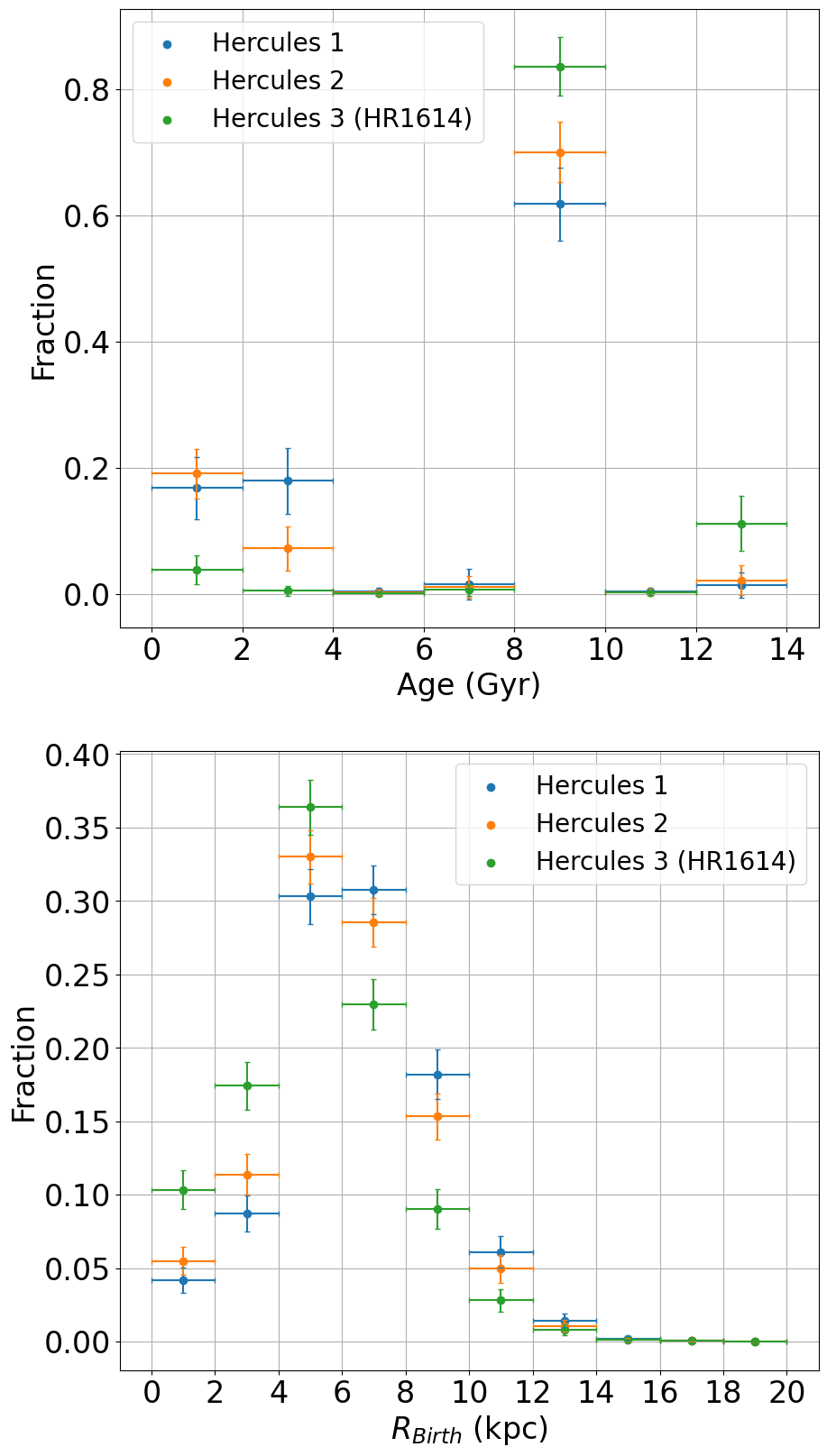}
	\caption{The fractional age (top panel) and birth radius (bottom panel) distributions for stars in the Hercules streams. All stars that fall within $2\sigma$ of a steam center (see Table \ref{tab:group_ellipses}) are used to calculate the fractional distributions. The total number of stars (and error on this total) in each age and birth radius bin is calculated based on the inferred fractions shown in the unequal area regions in Figures \ref{fig:age_dist} and \ref{fig:Rb_dist}.}
	\label{fig:hercules_dist}
\end{figure}

Specifically for the Hercules groups, \citet{torres2019} examined the age distribution of each branch using a sample of 12,227 white dwarfs and found that Hercules 1 and 2 \citep[a and b in][]{torres2019} had a more prominent peak at around 4 Gyr, and Hercules 3 \citep[c in][]{torres2019} was more uniform with slight peaks at around 3 Gyr and 8 Gyr. All groups from \citet{torres2019} displayed very few stars older than 10 Gyr. These results are consistent with what we find for Hercules 1 and 2 for the younger, more prominant peak, where we also find a peak in stars from $0-4$ Gyr (top panel; Figure \ref{fig:hercules_dist}). A difference in our distirubtion for Hercules 1 and 2 though is the prominent peak at $8-10$ Gyr that is present in both streams. For Hercules 3, we do find a more uniform distribution with the exception of a promient peak at 8$-$10 Gy, which is in a similar region as the lesser peak found in \citet{torres2019}.

Generally, the associations between age, metallicity and kinematics found here were also observed in \cite{wojno2018}. In \cite{wojno2018}, $\sim$12,000 FGK main-sequence turn-off stars with data from RAVE and Gaia DR1 were used to look at changes in kinematics based both on metallicity and age. In this 2018 study, older stars were found to be more kinematically hot and dispersed in the kinematic phase space, as are also now seen here. Also kinematic sub-structure was observed to change for bins in metallicity, demonstrating the possible chemical imprint of different potential features in the kinematics of stars in the Solar Neighborhood.

In a more detailed study, \citet{antoja2008} examined the change in the kinematic structure in the Solar Neighborhood as a function of age, similar to what has been done here, using a sample of $\sim$24,000 stars. They found that the major groups (i.e. Sirius, Coma Berenices, Hyades, Pleiades and Hercules) are detected over a larger range of ages, where Sirius, Coma Berenices and Hyades/Pleiades are detected at very young ages ($<2$ Gyr), and Hercules for ages $>$ 2 Gyr. Secondary peaks in the age distirbuion for Sirius, Coma Berenices and Hyades/Pleiades are also detected $\sim2-3$ Gyr. Additionally, the relative density of Hyades/Pleiades as compared to Hercules was found to decrease with increasing age. Except for the strong peak found for Sirius and Coma Berenices at very young ages ($<1$ Gyr), all of the main conclusions from \citet{antoja2008} are also supported in this study.

For the inferred birth radii distributions in Figure \ref{fig:Rb_dist}, we find clear substructure in the origin of these stars that seems to be correlated with the location of some of the kinematic groups. Using the metallicity of stars in these kinematic groups, \citet{bovy2010} found strong evidence that the Hyades group had a higher metallicity than expected, which suggests that it originates from regions inside the Solar circle. Additionally, they found weak evidence that the Sirius group had a lower metallicity than expected (and originated in the outer Galaxy) and that the Hercules group had a higher metallicity (and originated in the inner Galaxy). These results seem to be supported in Figure \ref{fig:Rb_dist}, where we see a slight excess of stars (as compared to grid regions with similar guiding radii) in the Hyades group at 4$-$6 kpc (as well as larger radii), a slight excess in Sirius at 10$-$12 kpc and an excess in Hercules 3 at 2$-$4 and 4$-$6 kpc. 

More recently, \citet{chiba2021} showed that due to resonances from the Galactic bar, the inner portion of the Hercules stream (corresponding to Hercules 3 in this work) is dominated by stars of small birth radii, and the outer portion of the Hercules stream (corresponding to Hercules 1 in this work) is dominated by stars of larger birth radii. Such a trend is supported by the inferred birth radii distributions in Figure \ref{fig:Rb_dist} and is shown more clearly by looking at the overall birth radius distribution for each of the streams (bottom panel of Figure \ref{fig:hercules_dist}). In Figure \ref{fig:hercules_dist}, we see that the mean birth radius shifts from lower to higher values as you go from Hercules 3 to 1, as predicted in \citet{chiba2021}. We also note that this change in birth radii seems well correlated with the inferred ages of the Hercules streams, as shown in the top panel of Figure \ref{fig:hercules_dist}. Here for the $8-10$ Gyr bin we see a decrease in the relative number of stars from Hercules 3 to 1 and see the opposite trend for the $2-4$ Gyr bin.

As our results seem to be supported by past studies, we now consider the new trends we uncover. First, we observe that not only do these kinematic groups differ from the background population of stars in some cases, as noted in past studies, we also find that there appears to be substructure in the age distributions \emph{within} these kinematic groups. Most notably, we see that in Sirius, Coma Berenices, Hyades and Pleiades, and Hercules 1 and 2, while most stars are in the 0$-$2 and 2$-$4 Gyr bins, the peak fraction in these bin does not always coincide with the kinematic center of group, as is most prominently the case for Coma Berenices. Additionally, we find that specific regions within some of these kinematic groups appear to be older than the general population within that group. For example, Coma Berenices has a clear enhancement of stars of 8$-$10 Gyr for stars with larger $U$ velocities, and Hercules 1 and 2 have some regions that have a much higher fraction of stars in the 8$-$10 Gyr range than the other streams discussed so far. As a final note, we also find here that Hercules 3 is the only stream that is almost entirely made of of stars of older ages ($>8$ Gyr).

The dominance of stars in the 0$-$2 and 2$-$4 Gyr bins for most streams, when combined with the metallicity distributions for these groups, also results in an inferred birth radii distribution that is dominant for the $6-8$ and 8$-$10 kpc bins. But again we see discrepancies \emph{within} some of the streams, notably for Coma Berenices. The excess of older 8$-$10 Gyr stars in Coma Berenices correlates with an excess of stars with birth radius 4$-$6 kpc. Additionally, the dominance of Hercules 3 stars at ages $>8$ Gyr correlates with a much smaller birth radii distribution peak and tail at smaller radii than the other streams.

This is especially interesting in regards to Coma Berenices, because multiple studies have concluded that the kinematic group demonstrates incomplete vertical phase mixing due to being predominantly present at negative Galactic latitudes \citep{monari2018, quillen2018b}. \citet{Mikkola2022} recently looked at the kinematic structure of white dwarfs in the Solar Neighborhood and found a distinct structure of faint white dwarfs at $(U, V) = (7, -19)$ km/s $-$ equivalent to $(U, x_{mix}) = (7, 7.9)$ in the present study. They suggest that if the fainter white dwarfs in this structure are older, then they could have a distinct dynamical origin. The age fractions in our Figure \ref{fig:age_dist} appear to corroborate this idea, because the part of Coma Berenices near the structure found in \citet{Mikkola2022} is exactly where we find a higher fraction of stars in the 8$-$10 Gyr bin. There is also clear evidence that this part of Coma Berenices may be an overlapping but different structure, as we also observe that the metallicity and dispersion in $W$ is quite distinct in this region (Figure \ref{fig:metal_W_UV_plane}). These findings seem to suggest a dynamical origin for this region that is distinct to the lower $U$ region that may not have undergone phase-mixing in the Galactic potential.

When looking at the overall distributions of the kinematic groups, we find that many of them seem to mostly be populated by stars from the most recent period of star formation (i.e. peaking 2$-$6 Gyr ago), while the background populations are dominated by stars from the first period of star formation in the disk (i.e. peaking 8$-$12 Gyr ago). The exception to this rule are for the Hercules streams, which, as a collection, seem to span the star formation history of the disk up until around 10 Gyr ago. This kind of insight could be helpful for future models that attempt to recreate the global potential in the Milky Way, as it gives some clues on the lifetime of the potential features that would cause these various kinematic structures.

\subsubsection{Evidence of Bending and Breathing Modes}\label{sec:breath_bend}

In the above we have focused on the substructure for the in-plane motion of these kinematic groups and how this relates to age and metallicity. From spectroscopic surveys over the past decade however it has been shown that there are also asymmetries in the motions perpendicular to the Galactic plane \citep{widrow2012, williams2013}. Specifically, when examining average $W$ velocity as a function of Galactic height above the mid plane, asymmetric relations (around the mid plane) with an odd parity are typically found: these are called ``breathing modes". Regions where a symmetric relation with even parity can also be identified: these are called ``bending modes". The density of stars is expected to be symmetric around the mid plane for breathing modes and asymmetric for bending modes. Following these observations, it was demonstrated that breathing modes can arise from asymmetric potentials within the Galaxy, like that of the bar and spiral arms \citep{faure2014, monari2015, monari2016}, while bending modes most likely arise from interactions with an external, Galactic satellite. Simulations have also shown that a passing satellite could produce both breathing and bending modes \citep{widrow2014}, as these external perturbers can themselves induce internal asymmetric potential features  \citep{widrow2015} that lead to breathing modes.

\begin{figure*}
	\includegraphics[width=\textwidth]{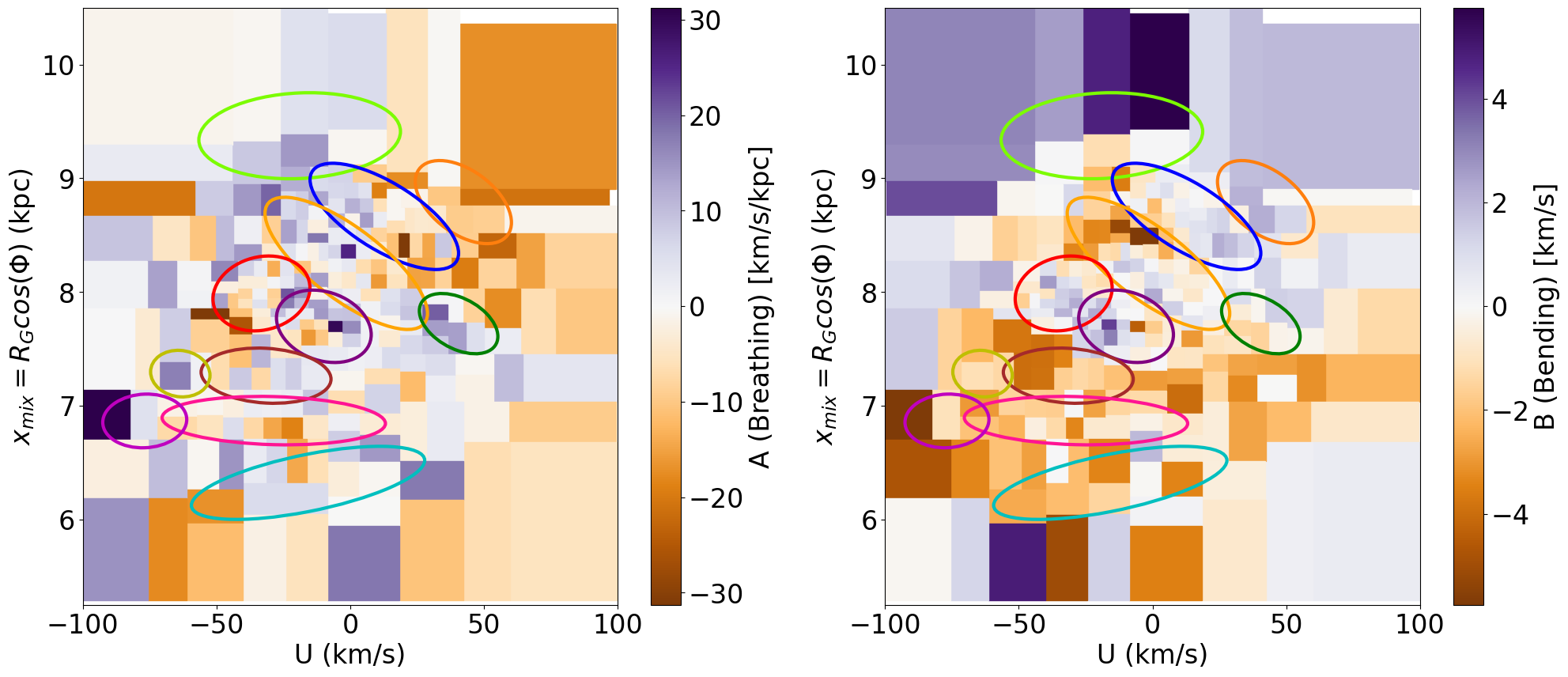}
	\caption{Maps of slope $A$ (left panel) and intercept $B$ (right panel), as described in eq. \ref{eq:bend_breath}, in the $x_{mix}$ vs. $U$ plane, where $A$ measures the relative strength of the breathing mode and $B$ the breathing mode. The unequal area regions contain an equal number of stars ($N = 971 \pm 1$) and are determined using the KD-Tree algorithm. The colored ellipses correspond to the kinematic groups identified in Section \ref{sec:kin_groups} and the colors correspond to the legend in Figure  \ref{fig:PCA_ellipses}.}
	\label{fig:bend_breath_params}
\end{figure*}

More recently, both bending and breathing modes have been observed with Gaia DR2 near the Sun \citep{gaiadr2_kinematics}, and we thus expect that such modes should be apparent in our data as well. To quantify the degree of each mode, we follow the convention from \citet{widrow2014} where for each subset of stars we fit:
\begin{equation}\label{eq:bend_breath}
	\overline{W}(z) = A \ z + B
\end{equation}
where $\overline{W}(z)$ is the median vertical velocity for some range of height above the Galactic mid plane, $z$, that is fit with a line of some slope $A$ and intercept $B$. For $\overline{W}(z)$, the median is bootstrapped over 1000 iterations, after correcting the vertical velocity assuming a peculiar velocity of the Sun is $W_\odot=7.25$ km/s \citep{Schonrich2010}, for 33 equal sized bins between $-$500 and 500 pc below/above the mid plane. Additionally, we only consider stars with $50^\circ < l < 200^\circ$, as we are incomplete for other ranges in Galactic longitude due to the requirement of Pan-STARRS photometry for determining metallicities. In eq. \ref{eq:bend_breath}, the slope $A$ measures the relative strength of the breathing mode, where a positive slope indicates an expanding breathing mode and a negative slope indicates a compressing breathing mode. The intercept $B$ on the other hand measures the relative strength of the bending mode, as a non-zero intercept will be fit for a group of stars whose density is asymmetric around the mid plane. For this study, we fit the linear relationship in eq. \ref{eq:bend_breath} to the stars in each of the irregular sized bins in Figures \ref{fig:age_dist} and \ref{fig:Rb_dist}. The best-fit values for the breathing and bending parameters $A$ and $B$ are shown in Figure \ref{fig:bend_breath_params}.

In Figure \ref{fig:bend_breath_params} we see some clear overall trends in the bending modes (right panel): there is an overall transition from positive to negative bending with decrease in $x_{mix}$. Such a trend has been observed in past studies \citep[e.g.][]{Schonrich2018, cheng2020, gaiaedr3_anticenter} and is consistent with the presence of a Galactic warp. From our Figure \ref{fig:bend_breath_params} however, we note significant deviations from the general trend, most evident in the Coma Berenices, Hyades and Pleiades kinematic groups. In low $U$ regions of Coma Berenices, we find much more negative bending modes than expected from the general trend for the warp, and in Hyades and Pleiades we find more positive bending modes than the Hercules groups directly below them. For Coma Berenices, these results seem to support the findings of incomplete phase mixing in this region \citep{monari2018, quillen2018b}, but this is still not consistent for larger values of $U$, supporting the findings in the previous section. 

Overall for the breathing parameter ($A$), we see less structured behavior, though globally it seems that stars within kinematic groups tend to experience expansion in their breathing behavior (i.e. positive slope).

A more interesting comparison is to examine the breathing and bending parameters for groups of stars of different ages. As we do not derive ages for individual stars in this study, we rely on regions in the $x_{mix}$ vs. $U$ plane that have stars predominantly for one age range. Additionally, even when a region is predominantly made up of one age range, we do not know which stars in the region belong to that age range. This is one of the major drawbacks of our chemodynamical age methodology. In an attempt to overcome this, we devise the following scheme for determining the breathing and bending parameters as a function of age.

In this scheme, we will find the values of $A$ and $B$ for regions in the $x_{mix}$ vs. $U$ plane as described above. The main difference here is for each region, we will determine a random age fraction for each age bin  from a normal distribution with mean and standard deviation equal to the values found from the above MCMC method. If this random age fraction is $>0.5$, then the majority of the stars in this region come from that age group and we will include the resulting $A$ and $B$ parameters in the following. We then select a random subset of stars from that region proportional to the random age fraction. With this random subset, we then calculate the $A$ and $B$ parameters as described above. This is then repeated for all regions in the $x_{mix}$ vs. $U$ plane, and the mean absolute $A$ and $B$ parameters are calculated for each age range using all regions that meet the random age fraction $>0.5$ criteria. We then bootstrap the mean and standard deviation on these parameters per age range for 1000 iterations. As a note, we also attempted this analysis when considering a random age fraction criteria of $>0.55$ and $>0.6$, and did not see a statistically significant change to the results presented here with the criteria of random age fraction $>0.5$.

\begin{figure}
	\centering
	\includegraphics[width=0.8\columnwidth]{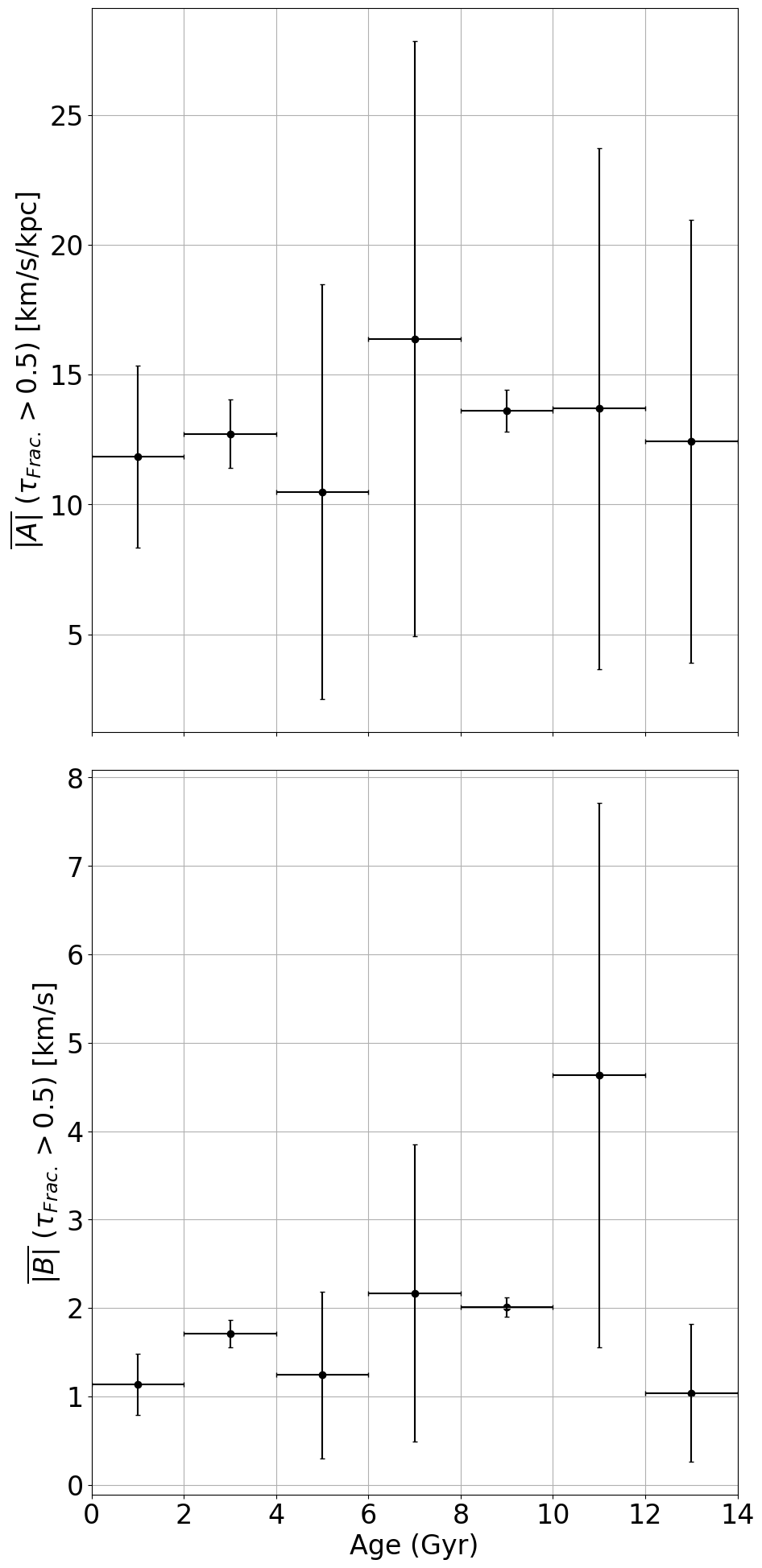}
	\caption{Mean absolute breathing parameter ($\overline{|A|}$; top panel) and absolute bending parameter ($\overline{|B|}$; bottom panel) for regions in the $x_{mix}$ vs. $U$ plane where the random age fraction is $>0.5$, as described in Section \ref{sec:breath_bend}. The means are bootstrapped for 1000 iterations and the resulting standard deviations are represented as error bars.}
	\label{fig:bend_breath_age}
\end{figure}

The resulting means, and standard deviations on the means, as a function of age are shown in Figure \ref{fig:bend_breath_age}. The first result from this is in the top panel of Figure \ref{fig:bend_breath_age}, where we 
seemingly observe no statistically significant trend in the breathing amplitude as a function of age. From the results in \citet{Ghosh2022} though, it was shown in both simulations and observationally with data from Gaia DR2 that the amplitude of breathing modes increases and then decreases again with the age of the population. The simulated galaxy in their work had predominant spirals that they showed led to the breathing modes, supporting the conclusion that the spiral arms induced the modes detected in the Gaia DR2 data. Observationally, \citet{Ghosh2022} found that their ``young" ([1, 4.85] Gyr) population had a lower absolute slope than the ``intermediate" ([4.85, 7.1] Gyr) population (though to a low level of statistical significance), and then that their ``old" ([7.1, 12.6] Gyr) population had a lower absolute slope than the intermediate one. Here we cannot confirm this trend. This demonstrates a major limitation of our age methodology. Primarily, because we cannot know which individual stars fall within a certain age range, this means we will have large errors for results in most analyses that try to rely/account for this. This is in contrast to overall population comparisons, like the one shown in Figure \ref{fig:hercules_dist}, which our method works well for.

When examining the trends for the bending mode strength as a function of age (Figure \ref{fig:bend_breath_age}, lower panel), we find hints that the bending mode strength slightly increases throughout the history of the disk, except for the oldest, thick disk stars. When comparing the youngest ($0-2$ Gyr) and oldest ($8-10$ Gyr) thin disk stars, we see a significant $\sim2.4\sigma$ increase in the bending amplitude. Between all other points though, we see $<2\sigma$ differences in the bending amplitude, so we cannot currently confirm this trend with the given data. Within the uncertainties though, we can at least conclude the bending amplitude is constant throughout the age of the disk, which is consistent with work from \cite{laporte2019}, where it was demonstrated with N-body simulations that interactions of the Milky Way with a Sagittarius-like dSph produced coupled vertical oscillations, like the ones seen here. These simulations produce a phase-space spiral in $v_z$ vs. $z$ that is observed with data from Gaia DR2 and demonstrated by \cite{laporte2019} to be persistent for stars with ages $>6$ Gyr. Our results seem to corroborate the presence of such a long-lived structure.

This is at odds with  results from \cite{wang2020}, who concluded that the warp amplitude decreased with age, which would indicate that the warp is induced by the nongravitational interactions from e.g., gas infall. This conclusion is based on the fact that as the youngest population should trace the gas in the Galaxy, it will always have the stronger bending amplitude in this scenario. Despite our relatively large errors, our results do disagree with this conclusion due to the significant increase in the bending amplitude between $0-2$ Gyr and $8-10$ Gyr.

In \cite{cheng2020}, it was found that the bending parameters were constant with age. However \cite{cheng2020} were only able to find bending parameters for two age ranges; $3-6$ Gyr and $6-9$ Gyr. For just these two ranges, we also find that the bending parameters are constant, as, again, the only significant difference for these parameters is when comparing our youngest age bin to the $8-10$ Gyr bin. Overall, our results best align with findings from \cite{romero2019}, who used OB stars, as a proxy for a young population, and red giant branch (RGB) stars, as a proxy for an older population, to show the warp is present in both populations, but that the amplitude increases with the age of the population, as seen here when comparing the $0-2$ Gyr and $8-10$ Gyr bins. This would suggest that our findings best align with the idea that the Galactic warp is the result of an external, gravitationally induced phenomenon. Because of the large uncertainties in the other bins though, there is still further analysis needed to confirm this.

Another interesting results is the bending amplitude for the range of $10-12$ Gyr, which seems to display a different heating history than the rest of the disk. Again, we note that there are still large uncertainties on this measurement. The age of this bin seems to align well with the merger event that may have led to the formation of the thick disk and the inner halo \citep{Belokurov2018, helmi2018}, commonly referred to as the Gaia-Enceladus-Sausage. It seems like the evidence of this merger may even be apparent in this local distribution of disk stars, where we see a more elevated bending amplitude for this specific age group of stars, though more work is needed to confirm this.

\section{Conclusions}\label{sec:conclusion}

In this study, we identified kinematic groups in the Solar Neighborhood based on the kinematics and Galactic location of low-mass stars. These kinematic groups are consistent with those found in studies of more massive main-sequence stars \citep[e.g.][]{ivezic2012, antoja2012, bovy2016, gaiadr2_kinematics, ramos2018, gaia_edr3_anticenter}. With these groups, we note clear differences in both metallicity and vertical velocity as compared to the surrounding regions in the $x_{mix}$ vs. $U$ plane, which we hypothesize is due to differences in mean age of the groups and field stars.

To confirm this and better understand these structures, we develop a method to estimate the probable age distribution of a group of stars based on their distribution in $W$ vs. metallicity. This method used Gaussian Mixture models of main-sequence turnoff stars from GALAH to define probability distributions in age bins of 2 Gyr. Using a MCMC method, we estimate the most probable contribution from each bin on the overall observed distribution to get a probable age distribution for the group. Using the GALAH subset, we validate this method for age distributions of various shapes and amplitudes and find we recover an accurate age distribution in most cases. Additionally, we use the resulting age distribution, in combination with the observed metallicity distribution of a group, to determine probable birth radii distributions for a group, which was based on work from \cite{frankel2020}.

With these methods, we estimate the probable age and birth radii distributions for groups of stars of equal number in the $x_{mix}$ vs. $U$ plane. Overall, the resulting age distributions are largely in agreement with what has been observed with main-sequence turnoff stars and white dwarfs in past studies \citep{antoja2008, wojno2018, torres2019}. The novelty is that the distributions examined here are based on a sample larger by order a magntiude, meaning variations can be observed at smaller velocity scales than in past studies. In general, we find that the regions associated with Sirius, Coma Berenices, Hyades and Pleiades groups are mostly represented by the most recent period of star formation (i.e. peaking 2-6 Gyr ago), while the background populations and one component of the Hercules stream have a more significant contribution from the first period of star formation in the disk (i.e. peaking 8-12 Gyr ago).These two peaks are consistent with the ``Two Infall Model" \citep{Chiappini1997} for the formation of the disk. As these former groups (i.e. Sirius, Coma Berenices, Hyades and Pleiades) are more commonly thought to be formed by spiral arms, this trend, and the finer age substructure in these regions, could help better constrain models of the Milky Way that consider the lifetime of these features. Additionally, we find an age gradient across the Hercules streams that is well correlated with peaks in birth radius, where such peaks in birth radius are also predicted in work by \citet{chiba2021}. Again, such a finding may help better constrain properties of the Galactic bar via modeling of these potential features.

Finally, we also examined the bending and breathing modes in the $x_{mix}$ vs. $U$ plane and how they relate to age. We find that the breathing modes do not show any statistically significant correlation with age. We also find that the bending amplitude is at the very least long-lived and shows hints of a slight increase with age, particularly when comparing the bending amplitudes in the $0-2$ Gyr and $8-10$ Gyr bins. This preliminary result best aligns with results from previous studies that conclude the warp was most likely induced by an external satellite \citep[e.g.][]{laporte2019, romero2019, cheng2020}, though additional work is needed to confirm this. We also observed a large, but not statistically significant, increase in the bending amplitude for stars of 10-12 Gyr in age, which seems to align well with the age of the thick disk that most likely formed from a distinct merger event \citep{Belokurov2018, helmi2018}. Similar to the breathing modes though, we are unsure about the significance of all trends due to large uncertainties on most age ranges, which demonstrates the main limitation of our age methodology in that analyses that require individual ages of stars may not benefit from the results from this paper.

In the future, an expanded sample of nearby low-mass stars with known metallicities and complete kinematics could allow us to conduct a similar study for (1) even smaller velocity scales and (2) varying radii and azimuth. This could potentially reveal even finer structures in age, and even show variation in these substructures with small changes in Galactic position. Future surveys, like SDSS-V, will provide spectra for 100,000s of low-mass stars in the Solar Neighborhood to enhance such a study. Overall though, the results and methodology laid out in this study has provided higher resolution chemodynamical age distributions in the Solar Neighborhood than before, which can better inform future models of dynamical interactions in our Galaxy.

\section*{Acknowledgments}

Mr.~Medan gratefully acknowledges support from a Georgia State University Second Century Initiative (2CI) Fellowship.

This work has made use of data from the European Space Agency (ESA) mission
{\it Gaia} (\url{https://www.cosmos.esa.int/gaia}), processed by the {\it Gaia}
Data Processing and Analysis Consortium (DPAC,
\url{https://www.cosmos.esa.int/web/gaia/dpac/consortium}). Funding for the DPAC
has been provided by national institutions, in particular the institutions
participating in the {\it Gaia} Multilateral Agreement.

This work has made use of data from Pan-STARRS. The Pan-STARRS Surveys (PS1) and the PS1 public science archive have been made possible through contributions by the Institute for Astronomy, the University of Hawaii, the Pan-STARRS Project Office, the Max-Planck Society and its participating institutes, the Max Planck Institute for Astronomy, Heidelberg and the Max Planck Institute for Extraterrestrial Physics, Garching, The Johns Hopkins University, Durham University, the University of Edinburgh, the Queen's University Belfast, the Harvard-Smithsonian Center for Astrophysics, the Las Cumbres Observatory Global Telescope Network Incorporated, the National Central University of Taiwan, the Space Telescope Science Institute, the National Aeronautics and Space Administration under Grant No. NNX08AR22G issued through the Planetary Science Division of the NASA Science Mission Directorate, the National Science Foundation Grant No. AST-1238877, the University of Maryland, Eotvos Lorand University (ELTE), the Los Alamos National Laboratory, and the Gordon and Betty Moore Foundation.

This work makes use of data products from the Two Micron All Sky Survey, which is a joint project of the University of Massachusetts and the Infrared Processing and Analysis Center/California Institute of Technology, funded by the National Aeronautics and Space Administration and the National Science Foundation.

This work makes use of data products from the Wide-field Infrared Survey Explorer, which is a joint project of the University of California, Los Angeles, and the Jet Propulsion Laboratory/California Institute of Technology, funded by the National Aeronautics and Space Administration.

This work made use of the Third Data Release of the GALAH Survey \citep{galahdr3}. The GALAH Survey is based on data acquired through the Australian Astronomical Observatory, under programs: A/2013B/13 (The GALAH pilot survey); A/2014A/25, A/2015A/19, A2017A/18 (The GALAH survey phase 1); A2018A/18 (Open clusters with HERMES); A2019A/1 (Hierarchical star formation in Ori OB1); A2019A/15 (The GALAH survey phase 2); A/2015B/19, A/2016A/22, A/2016B/10, A/2017B/16, A/2018B/15 (The HERMES-TESS program); and A/2015A/3, A/2015B/1, A/2015B/19, A/2016A/22, A/2016B/12, A/2017A/14 (The HERMES K2-follow-up program). We acknowledge the traditional owners of the land on which the AAT stands, the Gamilaraay people, and pay our respects to elders past and present. This paper includes data that has been provided by AAO Data Central (datacentral.org.au).

\section*{Data Availability}

All the data used in this paper is publicly available. The \textit{Gaia} data can be retrieved through the \textit{Gaia} archive (\url{https://gea.esac.esa.int/archive}), the 2MASS data can be retrieved via the NASA/IPAC Infrared Science Archive (\url{https://irsa.ipac.caltech.edu/Missions/2mass.html}), the Pan-STARRS data can be retrieved via MAST (\url{https://catalogs.mast.stsci.edu/panstarrs}), the AllWISE data can be retrieved via the NASA/IPAC Infrared Science Archive (\url{https://irsa.ipac.caltech.edu/Missions/wise.html}) and the GALAH data can be downloaded directly (\url{https://cloud.datacentral.org.au/teamdata/GALAH/public/GALAH_DR3/}). The photometric metallicties used in this paper are available as machine readable tables in the electronic version of the manuscript. The age and birth radii distributions, and the code to reproduce them can be found at: \url{https://github.com/imedan/chemo_dyn_ages}.

\bibliographystyle{mnras}
\bibliography{500_pc_chemodyamics_accepted.bib}

\end{document}